\documentclass[11pt]{article}
\usepackage{amsmath}
\usepackage{amssymb}
\usepackage{ntheorem}
\usepackage{graphicx}
\usepackage{subfigure}
\usepackage{color}
\usepackage{bm}
\usepackage{comment}

\def\YZ{{\bf Add refs}}

\numberwithin{equation}{section}
\newtheorem{theorem}{Theorem}[section]
\newtheorem{proposition}[theorem]{Proposition}
\newtheorem{definition}[theorem]{Definition}
\newtheorem{lemma}[theorem]{Lemma}
\newtheorem{remark}[theorem]{Remark}
\def\Proof{\noindent{\bf Proof} \quad}
\def\qed{\hfill $\Box$ \smallskip}

\def\N{{\mathbb N}}
\def\Z{{\mathbb Z}}

\def\R{{\mathbb R}}
\def\C{{\mathbb C}}

\def\bk{{\bf k}}

\def\mca{{\mathcal A}}

\def\mcc{{\mathcal C}}
\def\mce{{\mathcal E}}

\def\mcr{{\mathcal R}}

\def\mct{{\mathcal T}}

\def\mcp{{\mathcal P}}

\def\mcq{{\mathcal Q}}

\def\bx{{\bf x}}
\def\bx{{\bf x}}
\def\by{{\bf y}}
\def\bv{{\bf v}}

\def\bq{{\bf q}}

\def\buu{{\bf u}}

\def\c{{\cdot}}
\def\q{{\quad}}
\def\qq{{\qquad}}

\def\v{{\vert}}

\begin{document}

\title{Double Dirac Cone in Band Structures of Periodic Schr\"odinger Operators}

\author{Ying Cao\thanks{Yau Mathematical Sciences Center, Tsinghua Unversity, Beijing, 100084, China ({caoy20@mails.tsinghua.edu.cn}).}, \and Yi Zhu \thanks{Yau Mathematical Sciences Center, Tsinghua Unversity, Beijing, 100084, China, and Yanqi Lake Beijing Institute of Mathematical Sciences and Applications, Beijing, 101408, China({yizhu@tsinghua.edu.cn}).} }

\maketitle

\begin{abstract}
    Dirac cones are conical singularities that occur near the degenerate points in band structures. Such singularities result in enormous unusual phenomena of the corresponding physical systems. This work investigates double Dirac cones that occur in the vicinity of a fourfold degenerate point in the band structures of certain operators. It is known that such degeneracy originates in the symmetries of the Hamiltonian. We use two dimensional periodic Schr\"odinger operators with novel designed symmetries as our prototype. 
    First, we characterize admissible potentials, termed as super honeycomb lattice potentials. They are honeycomb lattices potentials with a key additional translation symmetry. It is rigorously justified that Schr\"odinger operators with such potentials almost guarantee the existence of double Dirac cones on the bands at the $\Gamma$ point, the origin of the Brillouin zone. We further show that the additional translation symmetry is an indispensable ingredient by a perturbation analysis. Indeed, the double cones disappear if the additional translation symmetry is broken. Many numerical simulations are provided, which agree well with our analysis. 
   \end{abstract}

{\small {\bf MSCcodes}:
35Q40, 35Q60, 35P99

}

\tableofcontents\newpage

\section{Introduction} \label{intr}
The Dirac cone is a conical structure near the degenerate point on the energy bands. It is deeply rooted in the symmetries of operators. Single Dirac cones near a twofold degenerate point have been found in many physical systems. It is a hallmark and reveals the underlying mechanism of versatile electronic or photonic properties of topological materials \cite{RN98,RN99,RN101,RN103,RN104}. A typical system possessing the single Dirac cone is the two-dimensional material--graphene, which has the atomic honeycomb lattice made of carbon atoms \cite{RN91,RN13,RN14}. Its great success in many fields has brought the blooming time for both experimental and theoretical understanding of such degenerate points on spectral bands. Meanwhile, other types of conically degenerate points were reported. Among those, double Dirac cones have attracted considerable attention \cite{RN61, RN62}. Such conical structures consist of two cones that share a fourfold degenerate apex. Due to the higher degeneracy, the corresponding wave patterns and physical properties are different from those in systems possessing single Dirac cones \cite{RN102,RN106, RN73,RN74}. It is known that the nontrivial topology of energy bands and corresponding significant properties of materials are born from the singularity. Thus, investigating the underlying symmetries and degeneracy related to the double Dirac cone will help us better understand the unusual physical properties.

Regarding single Dirac cones, a lot of analyses about the related time reversal symmetric operators have been done through different models and vehicles, especially when the material has a honeycomb structure. The tight-binding approximation was first developed by Wallace to describe the band structure of graphite \cite{RN87}, and later used systematically by others \cite{RN88, RN89}. The perturbation theory and multiscale analysis help to solve shallow potential cases successfully \cite{RN94,RN95}. One pioneering rigorous result on characterizing the honeycomb potentials and demonstrations of the existence of Dirac points was given by Fefferman and Weinstein \cite{RN18}. They paved the way to rigorously analyzing such degenerate spectral points by combining Lyapunov-Schmidt reduction, perturbation theories, and multidimensional complex analysis. Based on their results, many other problems were solved such as the evolution of wave packets spectrally concentrated near Dirac points \cite{RN9}, edge states and valley Hall effect \cite{RN5}, lower dimensional degenerate points \cite{RN84} and threefold Weyl points in the three-dimensional problems \cite{RN77}. Ammari and collaborators did a lot of work on the Dirac cone and edge states using layer potential theory in the subwavelength regime \cite{RN96, RN113, Ammari2020}. Besides, the group representation theory has been used by Berkolaiko and Comech to describe the symmetric structure \cite{RN97}. Despite the aforementioned progress in this blooming area, there is rarely any rigorous result on double Dirac cones as those in single Dirac cones.  

In this paper, we investigate the two dimensional Schr\"odinger operator $H_V = -\Delta + V(\bx)$ with $V(\bx)$ specially structured such that $H_V$ has a double Dirac cone on its energy surfaces. Our goal is to find the precise mathematical description of this special kind of $V(\bx)$, and establish the rigorous proof about the existence of the double Dirac cone. We first define a class of potentials $V(\bx)$, termed as the super honeycomb lattice potentials. They are honeycomb lattice potentials equipped with an additional translation symmetry. Then we prove that such $V(\bx)$ is enough for the existence of a double Dirac cone at $\Gamma$ point, the origin of the Brillouin zone, as is stated in the main theorem Theorem \ref{thm-main}. To achieve our goal, we utilize Lyapunov-Schmidt reduction, perturbation theory, and spectral theories about infinite dimensional linear operators. The rigorous analysis is inspired by pioneering works on single Dirac cones by Fefferman and Weinstein \cite{RN18}. However, due to higher multiplicity and the additional symmetry, we need a more delicate decomposition of the working function spaces and the bifurcation matrix, see Sections \ref{function spaces} and \ref{double-cones-after-degeneracy}. We also show that the extra translation symmetry is indispensable. Namely, a small perturbation that breaks this symmetry leads to the separation of the fourfold degeneracy and disappearance of the double cone in the band structure, see Section \ref{Perturbations}. Besides, we give two typical examples of potentials that are in the class of admissible potentials. Numerical simulations are provided to support our analysis. Our results will shine a light on the study of more complicated symmetries of operators and higher degeneracy on energy bands.

The rest of the paper is organized as follows. Section \ref{super-honeycomb} provides the preliminaries. The definition of super honeycomb lattice potentials and the decomposition of the working function space are given based on symmetries. In Section \ref{Double-cone}, we state and prove the main theorem--the existence of the double Dirac cone at the $\Gamma$ point of the Schr\"odinger operator with a super honeycomb lattice. Inspired by \cite{RN18}, the proof is divided into two main parts. First, we show that the fourfold degeneracy at the $\Gamma$ point leads to a double cone in the vicinity under proper assumptions. Secondly, we justify the assumptions for shallow potentials and then extend the shallow potentials to generic potentials. In Section \ref{Perturbations}, we discuss the band structures under perturbations which break the additional translation symmetry.  The double Dirac cone separates into two parts and a local energy gap appears near the $\Gamma$ point. At the end, corresponding numerical simulations for the two typical potentials are {given} in Section \ref{Num-Res}.

\section{Super honeycomb lattice potential and symmetries}\label{super-honeycomb}

Symmetries of an operator are the origin of many novel properties of its spectrum. In this section, we introduce a large class of potentials, termed as super honeycomb lattice potentials, which are characterized by several symmetries. Their properties and corresponding spectral theory are discussed.

\subsection{Super honeycomb lattice potentials}
We first introduce the parity, complex-conjugation, and rotation operators for a function $f(x)$ defined in $\R^2$ as below:
$$\mcp[f](\bx) = f(-\bx), \quad \mcc [f](\bx) = \overline{f(\bx)}, \quad \mcr_\theta[f](\bx) = f(R_\theta^*\bx),$$
where 
\begin{equation}\label{eqn-rotation-matrix-clock}
    R_\theta = \begin{pmatrix}
    \cos{\theta} & \sin{\theta} \\
    -\sin{\theta} & \cos{\theta}
    \end{pmatrix}
\end{equation}
represents the clockwise rotation by an angle of $\theta \in [0, 2\pi]$ in $\R^2$ and its Hermitian $R_\theta^*=R_\theta^{-1}$ represents the anticlockwise rotation. A function $f(x)$ is called $\mcp-$invariant (or parity symmetric) if $\mcp[f](\bx) = f(\bx)$, and similar for $\mcc-$ invariant (or conjugation symmetric) and ${\mcr_{\theta}}-$invariant (or rotation symmetric). Besides, being $\mcp\mcc$ invariant is also called having the time reversal symmetry. In this work, we are interested in $R_{\frac{2}{3}\pi}$, so we omit the subscript, i.e., $R = R_{\frac{2}{3}\pi} $ and $\mcr = \mcr_{\frac{2}{3}\pi} $ for simplicity.

We also use the following notation for the translation operator of a nonzero vector $\buu$ in this article:
$$\mct_{\buu} [f] (\bx) = f(\bx + \buu).$$

In this work we are interested in the spectra of $H_V=-\Delta + V(\bx)$ with the potential $V(\bx)$ being equipped with the above symmetries. Namely, we have the following definitions.  
\begin{definition}\label{def-honeycomb}
    $V(\bx)\in L^{\infty}(\R^2)$ is called a honeycomb lattice potential, if
\begin{enumerate} 
    \item $V(\bx) $ is real and even,
    \item $V(\bx)$ is $\mcr$-invariant,
    \item $V(\bx)$ has a period $\buu_1\neq 0$, and thus $\buu_2 - periodic$ with $\buu_2 = -R^*\buu_1$.
\end{enumerate}
\end{definition}

\begin{remark}
    {
        All these three properties are discussed in $L^{\infty}(\R^2)$, that is to say, $V(\bx) = \mcc V(\bx) = \mcp V(\bx) = \mcr V(\bx) = \mct_{\buu_1} V(\bx) = \mct_{\buu_2}V(\bx )$ is valid almost everywhere. The following discussion about super honeycomb lattice potential is also in $L^{\infty}(\R^2)$ in the same way.}
\end{remark}

\begin{figure}[htbp]
    \centering
    \subfigure[]{\includegraphics[width=6cm]{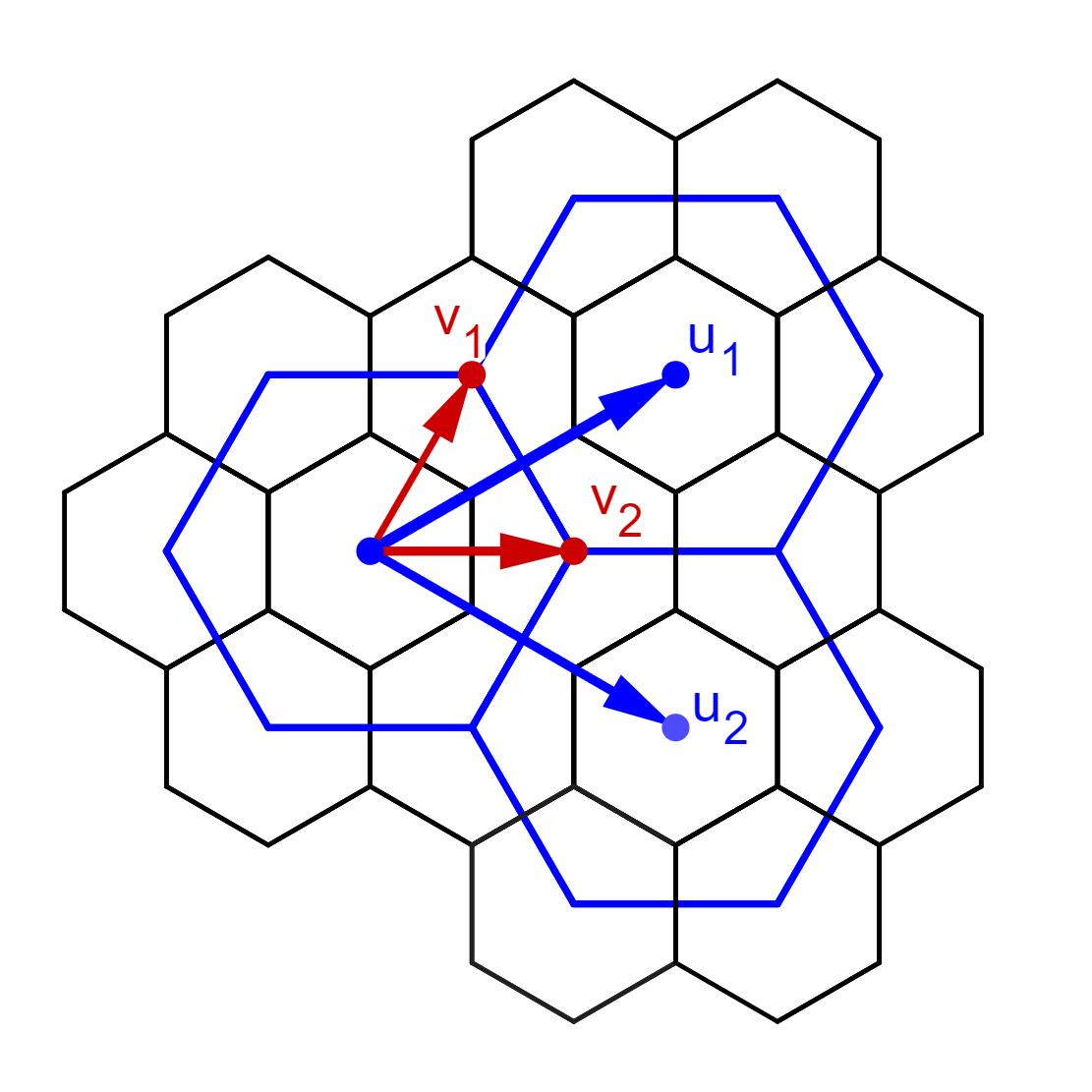}}
    \subfigure[]{\includegraphics[width = 6cm]{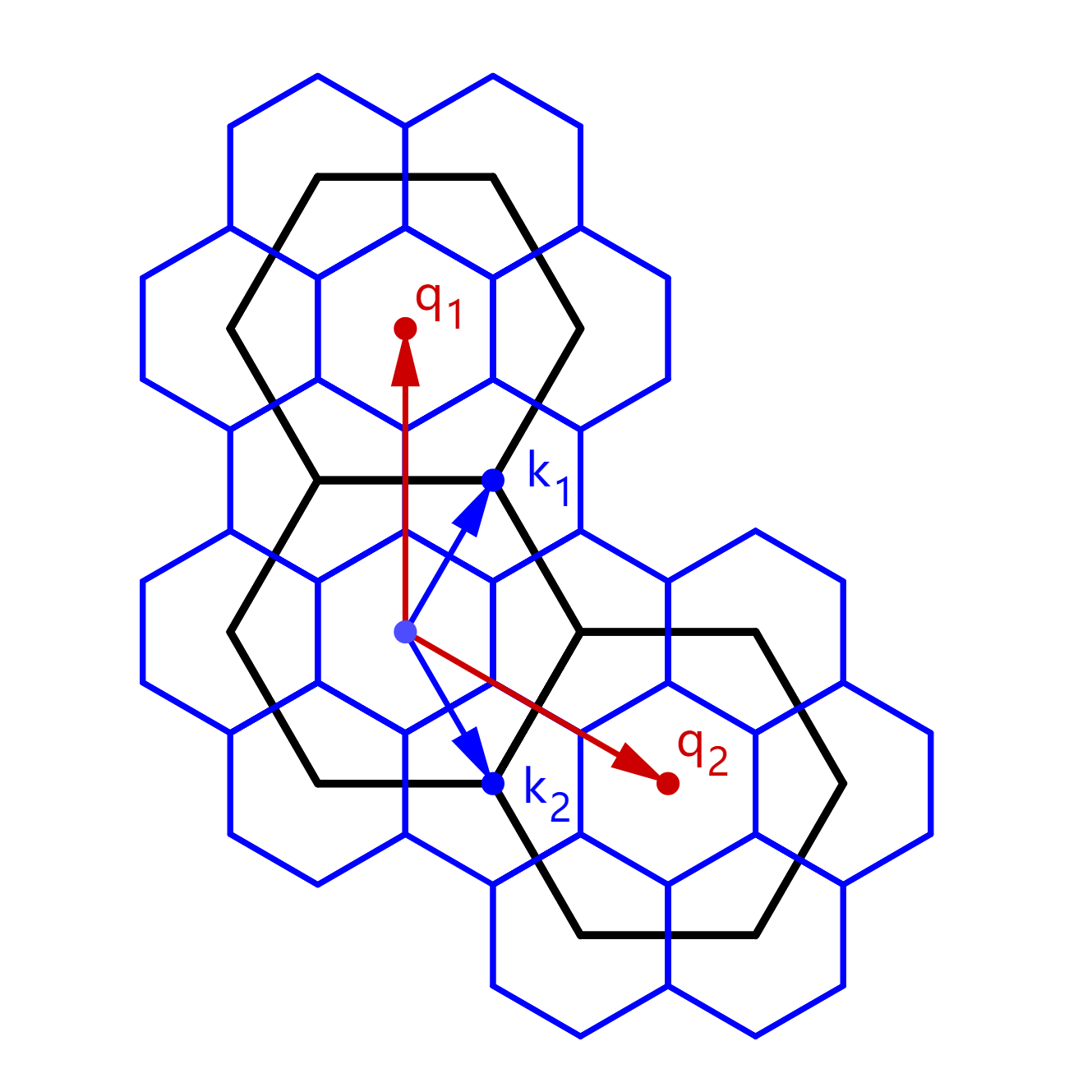}}
    \caption{(a) the figure of lattice, and (b) the figure of dual lattice. The blue lattices are the cases of honeycomb lattice, with periods $\buu_1$ and $\buu_2$ and dual periods $\bk_1$ and $\bk_2$.
     The black lattices are the cases of corresponding super honeycomb lattice, with periods $\bv_1$ and $\bv_2$ and dual periods $\bq_1$ and $\bq_2$. Obviously, the super honeycomb lattice is a kind of honeycomb lattice, but not vice versa.}
    \label{fig-lattice}
\end{figure}

Therefore, the non-relativistic Sch\"odinger operator $H_V(\bx) = -\Delta + V(\bx)$ has time reversal symmetry, rotation symmetry, and translation symmetry if $V(\bx)$ is a honeycomb lattice potential.
The honeycomb lattice can refer to the blue lattice in Figure \ref{fig-lattice}. 
As shown in this Figure, the black lattice is a honeycomb lattice, too. But it has an extra translation symmetry with periods $\bv_1$ and $\bv_2$, where
\begin{equation}\label{eqn-bv}
    \bv_1 = \frac{1}{3}(2\buu_1 - \buu_2), \qq \bv_2=\frac{1}{3}(\buu_1+\buu_2).
\end{equation}
We call it a super honeycomb lattice because of this additional symmetry. 

\begin{definition}\label{def-superHoneycomb}
    A honeycomb lattice potential $V(\bx)\in L^{\infty}(\R^2)$ is called a super honeycomb potential if 
    \begin{enumerate}
        \item[4.] V(\bx) is $\bv_1$ and $\bv_2$ periodic,
        where $\bv_1$ and $\bv_2$ are as in (\ref{eqn-bv}) and the non-degeneracy condition holds:
        \begin{equation}\label{eqn-nondegeneracy}
           \frac{1}{|\Omega|}\int_{\Omega} e^{-i \bq_1 \cdot \by}V(\by) d \by \neq 0 ,
        \end{equation}
        where $\Omega$ is a unit cell of honeycomb lattice as in (\ref{eqn-unit-cell}). 
    \end{enumerate}
\end{definition}

By definitions, a super honeycomb lattice potential is a honeycomb lattice potential with an additional translation symmetry. In other words, it has a smaller lattice structure. Since in applications we need to break this symmetries to obtain bands with different topological indices \cite{RN108,RN106}, see also Figure 2. We still consider the super honeycomb lattice potential in the bigger lattice structure.  

We remark that the non-degeneracy condition (\ref{eqn-nondegeneracy}) in the definition ensures the lowest Fourier coefficients of $V(\bx)$ do not vanish.
As a consequence, the degeneracy occurs at $2^{nd}-7^{th}$ bands. While higher bands should be considered if the non-degeneracy condition does not hold, which will be investigated in future works.

Here we give a typical example of super honeycomb lattice potentials which is a dimerization of a honeycomb lattice potential. It is a more general mathematical construction of the case studied in \cite{RN61}. 
 Our approach is first dimerizing in one direction, and then applying rotations to get the final results.
Beginning from a honeycomb lattice potential, there should be three directions for dimerization: $\buu_1$, $\buu_2$, and $\buu_3 = \buu_2-\buu_1$.
Thus, the following steps are needed to construct the dimer model.

Assume that $f(\bx)$ is a function such that $f(\bx+\frac{1}{2}\buu_3)$ is a honeycomb lattice potential, to be specific:
\begin{equation}\label{eqn-dimer-f}
    f(\bx+\buu_1) = f(\bx),\q f(\bx+\buu_2) = f(\bx), \q  \mcr f(\bx+\frac{1}{2}\buu_3) =  f(\bx+\frac{1}{2}\buu_3),\q \forall x \in \R^2.
\end{equation}
First dimerize in the $\buu_3$ direction:
\begin{equation}\label{eqn-dimer-g}
    g(\bx, r) = f(\bx - \frac{1}{2}r\buu_3) + f(\bx + \frac{1}{2}r\buu_3).
\end{equation}
Here $r\in[0,1]$ is the distance ratio for dimers.
Then rotates the obtained $g(\bx)$ :
\begin{equation}\label{eqn-dimer-W}
    W(\bx, r) = g(\bx,r) + \mcr g(\bx,r) + \mcr^2 g(\bx,r).
\end{equation}
$W(\bx,r)$ is the dimer model we want. Using (\ref{eqn-dimer-f})-(\ref{eqn-dimer-W}), it is easy to check 
the conclusion below.

\begin{proposition}
    Any $W(\bx, \frac{1}{3})$ constructed by the above steps is a super honeycomb lattice potential.
\end{proposition}

\begin{figure}[htbp]
   \centering
   \subfigure[]{\includegraphics[width=3.8cm]{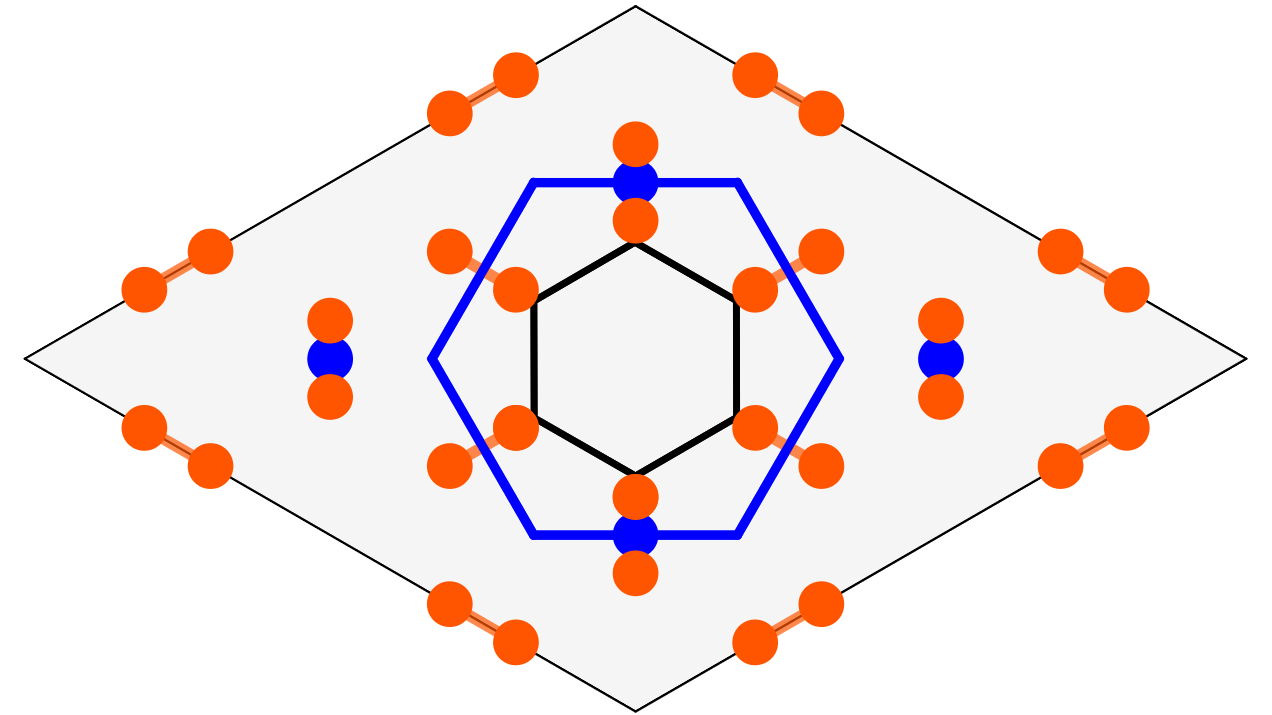}}\q
   \subfigure[]{\includegraphics[width=3.8cm]{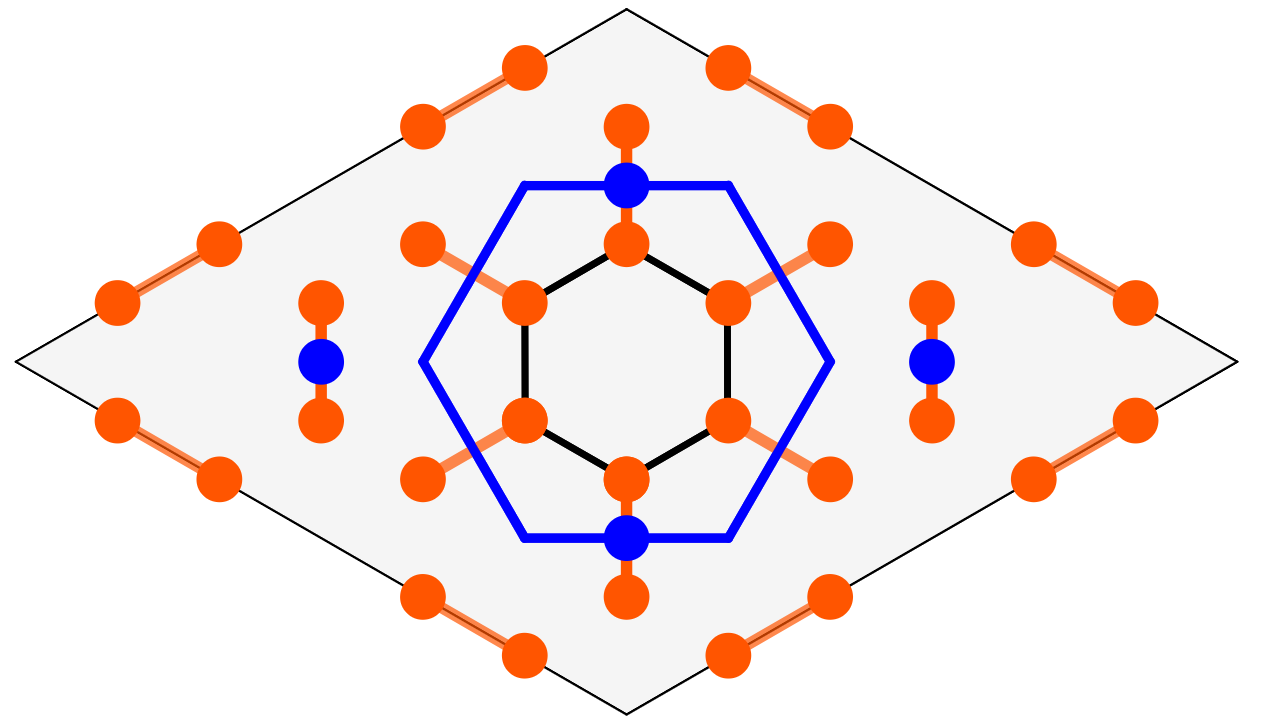}}\q
   \subfigure[]{\includegraphics[width=3.8cm]{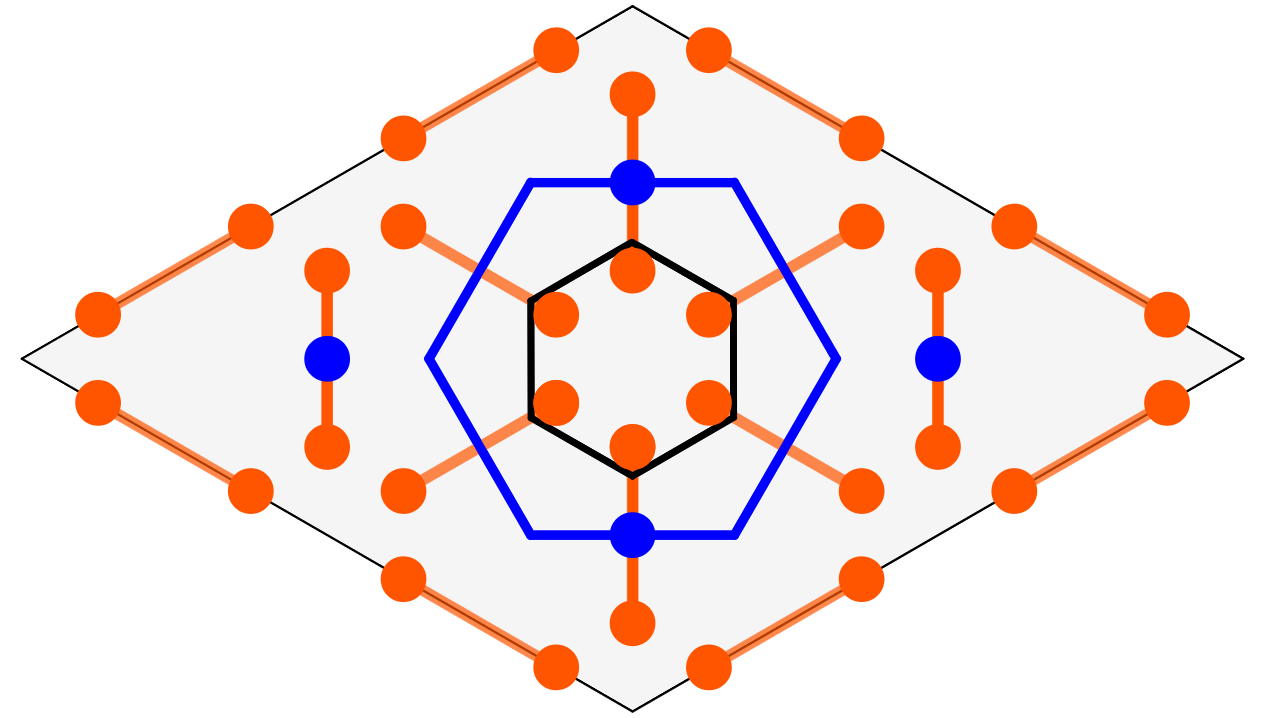}}
   \caption{Discrete honeycomb-dimer model examples with $r$ taking values (a) $\frac{13}{60}$, (b) $\frac{1}{3}$, and (c) $\frac{1}{2}$. The blue hexagon is a unit cell of honeycomb lattice, and the black hexagon is a unit cell of the super honeycomb lattice.
   Blue points are elements before dimerization and rotation, red points are dimers. Elements in a dimer are connected by a red line.} \label{fig-dimer}
\end{figure}

Figure \ref{fig-dimer} shows a discrete example.

Before all the detailed analysis of the energy surfaces of Schr\"odinger operators with super honeycomb lattice potentials, we try to first explain our method using symmetries.
If $V(\bx)$ is a super honeycomb lattice potential, the operator $H_V$ has following properties.

\begin{lemma}\label{lem-symmetry}
    Assume that $V(\bx)$ is a super honeycomb lattice potential, then  
    $${[H_V, \mct_{\bv_1}] =[H_V, \mct_{\bv_2}]= [H_V, \mcr]= 0}
    $$
    $$[H_V, \mcp] = [H_V, \mcc] = [H_V, \mcp\mcc]= 0 $$
    \end{lemma}
    \Proof Take $[H_V,\mcp]=0$ as an example and others are the same. For any $\phi(\bx)$,  
    $$H_V\mcp \phi(\bx)=(-\Delta + V(\bx)) \phi(-\bx) =  (-\Delta_{-x} + V(-\bx))\phi(-\bx) = \mcp H_V \phi(\bx),
    $$
    because $V(\bx)$ is even, and the Laplace operator is rotation invariant. 
    \qed
     
    {
    The spectrum of $H_V$ on $L^2(\R^2)$ is equivalent to such a union by Floquet-Bloch theorem:
    \begin{equation*}
        \sigma_{L^2(\R ^2)}(H_V) = \bigcup _{\bk \in \Omega^*}\sigma_{\chi}(H_V(\bk)),
    \end{equation*}
    where the notations are introduced in the following two subsections. $\chi$ is as in (\ref{eqn-chi-space}). Thus, finding eigenvalues of $H_V$ on $L^2(\R^2)$ is transformed into finding eigenvalues of $H_V(\bk)$ on $\chi$. It is clear that $H_V(\mathbf{0}) = H_V$ - the operator corresponding to $\Gamma$ point are commutative with $\mct_{\bv_1}$ and $\mcr$. $\mct_{\bv_1}$ and $\mcr$ are unitary and have three eigen-subspaces corresponding to different eigenvalues $\xi_1 = 1$, $\xi_2=e^{\frac{2}{3}\pi i}$, and $\xi_3=e^{- \frac{2}{3}\pi i}$ on $\chi$. Suppose $\phi_l(\bx)$ and $\phi_j(\bx)$ are any normalized eigenfunctions of $\mcr$ with eigenvalues $\xi_l$ and $\xi_j$, then:
    \begin{equation*}
    \begin{aligned}
        \langle H_V \phi_l(\bx), \phi_j(\bx) \rangle & = \langle \mcr H_V \phi_l(\bx), \mcr \phi_j(\bx) \rangle \\
        & = \overline{\xi_l}\xi_j \langle H_V \phi_l(\bx), \phi_j(\bx) \rangle = \delta_{l,j}\langle H_V \phi_l(\bx), \phi_j(\bx) \rangle, 
    \end{aligned}
    \end{equation*}
    where the inner product is as in (\ref{eqn-innerPro}). This tells that eigen-subspaces of $\mcr$ are invaraint spaces of $H_V$, and similar for $\mct_{\bv_1}$. Thus, we can further decompose the spectrum of $H_V$ on $\chi$ to the spectrum of $H_V$ on these subspaces. Finally, we associate these subspaces by $\mcp$, $\mcc$, and $\mcp\mcc$ and construct the degeneracy by the fact that they are commutative with $H_V$.}

\subsection{A quick review of Floquet-Bloch theory }\label{Periodic-potential}

In this subsection, we review the Floquet-Bloch theory briefly. Let $\{\buu _1,\buu _2\}$ be linear independent vectors in $\R^2$. Its corresponding equilateral lattice is
$\bf{U} = \Z \bf{u} _1 \oplus \Z \bf{u} _2$. Denote a unit cell 
\begin{equation}\label{eqn-unit-cell}
    \Omega = \{\buu = c_1\buu_1 + c_2\buu_2,\q c_1,c_2\in [0,1]\}.
\end{equation}
 Dual lattice of $\bf{U}$ is
$${\bf{U}}^* = \Z{\bk}_1 \oplus \Z {\bk}_2 \q {\buu}_l\cdot\bk_j = 2\pi \delta_{l,j}  \q l,j=1,2.$$ $\Omega^* = \R^2/\bf{U}^{*} $ is the Brillouin zone. We can divide the eigenvalue problem of $H_V$ on $L^2(\R^2)$ into the following eigenvalue problems of $H_V$ traversing all the $\bk$ in the Brillouin zone \cite{Spectral1973,Kemp1959}.

For $\bk \in \Omega^*$, consider the eigenvalue problem
    \begin{equation} \label   {eqn-Hv-eigen-problem-1}
        H_V \Phi(\bx,\bk) = \mu(\bk) \Phi(\bx,\bk) , \q \bx \in \R^2,
\end{equation}
\begin{equation} \label{eqn-Hv-eigen-problem-2}
        \Phi(\bx+\buu, \bk) = e^{i\bk \cdot\buu  }\Phi(\bx,\bk),\q  \buu \in \bf{U} ,\q \bx \in \R^2
\end{equation}

Let $\Phi(\bx ,\bk) = e^{i \bk \cdot \bx} \phi(\bx, \bk) $, where $\phi(\bx)$ is periodic. Then (\ref{eqn-Hv-eigen-problem-1}) is equal to:
$$
    (-\Delta + V(\bx)) (e^{i \bk \cdot \bx} \phi(\bx,\bk))= \mu e^{i \bk \cdot \bx} \phi(\bx,\bk)$$
that is
$$(-(\nabla + i\bk)\cdot(\nabla + i\bk) + V(\bx)) \phi(\bx,\bk) = \mu \phi(\bx,\bk) \q.$$
Let $H_V({\bk}) = -(\nabla + i\bk)\cdot(\nabla + i\bk) + V(\bx)$. Thus,  the eigenvalue problem can be rewritten as:
\begin{equation}\label{eqn-Hvk-eigen-problem-1}
    H_V({\bk}) \phi(\bx,\bk) = \mu \phi(\bx,\bk),\q \bx \in \R^2
\end{equation}
\begin{equation}\label{eqn-Hvk-eigen-problem-2}
    \phi(\bx+\buu, \bk) = \phi(\bx),\q  \buu \in \bf{U} 
\end{equation}

(\ref{eqn-Hv-eigen-problem-1})-(\ref{eqn-Hv-eigen-problem-2}), or equivalently (\ref{eqn-Hvk-eigen-problem-1})-(\ref{eqn-Hvk-eigen-problem-2}), has a real, discrete and lower bounded spectrum \cite{evans2010}:
$$
\mu_1(\bk) \leq \mu_2(\bk) \leq \mu_3(\bk) \leq...
$$

These energy bands have the following property \cite{RN18}.
\begin{lemma}\label{lem-mu-lipschitz}
Any $\mu_b(\bk)$ is periodic and Lipschitz continuous. Its periods are $\bk_1$ and $\bk_2$.
\end{lemma}

\subsection{Function spaces and symmetries at $\Gamma$ point}\label{function spaces}
For honeycomb lattice potentials, let $\buu_2 = -R \buu_1$. We concern about the spectrum at $\Gamma$ point most, so define
\begin{equation}\label{eqn-chi-space}
    \chi = \{ f(\bx) \in L^{2}_{loc}(\R^2), \q f(\bx+\buu) = f(\bx), \q \forall \buu \in \bf{U} \}.
\end{equation}
$\chi$ is a Hilbert space under the inner product:
\begin{equation}
    \langle f(\bx), g(\bx) \rangle = \frac{1}{|\Omega|}\int_{\Omega} \overline{f(\bx)} g(\bx) d \bx
\end{equation}
Our aim is to find the fourfold degeneracy of $H_V$ on $\chi$ and the double Dirac cone in the vicinity of this highly degenerate point $\Gamma$. We also define the limitation of $\chi$ in $H^1_{loc}(\R^2)$: 
\begin{equation}
    H^1_{per} = \{ f(\bx) \in H^1_{loc}(\R^2), \q f(\bx+\buu) = f(\bx), \q \forall \buu \in \bf{U} \}.
\end{equation}

A super honeycomb lattice potential $V(x)$ is not only real and in $\chi$, but also in 
\begin{equation}\label{eqn-small-period-space}
    \chi_s = \{f \in L^{2}_{loc}(\R^2)  ,\q f(\bx+\bv) = f(\bx), \q \bx\in \R^2,\bv \in \Z\bv_1 \oplus \Z\bv_2 \},
\end{equation}
where $\bv_1$ and $\bv_2$ are in the form of (\ref{eqn-bv}).
It is easy to see that 
\begin{equation}
    \buu_1 = \bv_1 + \bv_2, \qq \buu_2 = -\bv_1 + 2\bv_2,
\end{equation}
which means $\chi_s \subset \chi$, or functions in $\chi_s$ have smaller periods than those in $\chi$, as we have mentioned before.
Dual vectors for $\{\bv_1, \bv_2 \}$ such that $\bv_i \bq_j = \delta_{i,j}$ are:
\begin{equation}\label{eqn-small-dualperiod}
    \bq_1 = \bk_1-\bk_2 , \qq
    \bq_2 = \bk_1+ 2\bk_2.
\end{equation}

First, we claim a decomposition of $\chi$. This decomposition associates the extra translation symmetry with parity symmetry perfectly.

\begin{proposition}\label{prop-space-decomposition-wrt-small-period}
$\chi =\chi_{s} \oplus \chi_{\bk_1} \oplus \chi_{-\bk_1}$, where

\begin{equation}\label{eqn-exp(ik_1x)-small-period-space}
    \chi_{\bk_1} = \{f(\bx) = e^{i\bk_1\cdot \bx} p(\bx) ,\q p(\bx) \in \chi_s \},
\end{equation}
\begin{equation}\label{eqn-exp(-ik_1x)-small-period-space}
    \chi_{-\bk_1} = \{f(\bx)= e^{-i\bk_1\cdot \bx} p(\bx) ,\q p(\bx) \in \chi_s \}.
\end{equation}
And $\chi_s$, $\chi_{\bk_1}$ and $\chi_{-\bk_1}$ are eigen-spaces of $\mct_{\bv_1}$ with eigenvalues $1$, $e^{-\frac{2}{3}\pi i}$ and $e^{\frac{2}{3}\pi i}$.
\end{proposition}

\Proof
It is obvious that  $\chi_s$, $\chi_{\bk_1}$, and $\chi_{-\bk_1}$ are subsets of $\chi$. Also, it is easy to verify they are eigenspaces after some very simple calculations.
Thus, only need to prove that they are orthogonal to each other and $\chi \subset \chi_s \oplus \chi_{\bk_1}\oplus \chi_{-\bk_1}$.

According to Fourier analysis, 
\begin{equation}\label{eqn-fourier-large-period}
    \{ e^{i(m_1\bk_1+m_2\bk_2)\c \bx} \}_{(m_1,m_2)\in\Z^2}
\end{equation}
forms a Hilbert basis of $\chi$. Similarly, 
\begin{equation}\label{eqn-fourier-small-period}
    \chi_s = \overline{span \{ e^{i(n_1\bq_1+n_2\bq_2)\c \bx}, \q (n_1,n_2)\in \Z^2 \}  }
\end{equation}
From (\ref{eqn-small-dualperiod}), (\ref{eqn-fourier-small-period}) is equivalent to
\begin{equation}\label{eqn-fourier-L^0}
\chi_s = \overline{span \{ e^{i((n_1+n_2)\bk_1+(2n_2-n_1)\bk_2)\c \bx}, \q (n_1,n_2)\in \Z^2 \}  }
\end{equation}
And therefore
\begin{equation}\label{eqn-fourier-L^+}
    \chi_{\bk_1} = \overline{span \{ e^{i((n_1+n_2+1)\bk_1+(2n_2-n_1)\bk_2)\c \bx}, \q (n_1,n_2)\in \Z^2 \}  }
\end{equation}
\begin{equation}\label{eqn-fourier-L^-}
   \chi_{-\bk_1} = \overline{span \{ e^{i((n_1+n_2-1)\bk_1+(2n_2-n_1)\bk_2)\c \bx}, \q (n_1,n_2)\in \Z^2 \}  }
\end{equation}

The $\Z^2$ solutions of equations 
$$
    \left\{
    \begin{aligned}
    n_1 + n_2 = m_1 \\
    2n_2-n_1 = m_2
    \end{aligned}
    \right.
$$
exists: $(n_1, n_2) = (\frac{2m_1-m_2}{3},\frac{m_1+m_2}{3} )$, which means $e^{i(m_1\bk_1+m_2\bk_2)\c \bx}$ is in $\chi_s$, if and only if $(m_1+m_2)\equiv 0 \pmod 3$. It is easy to check, similarly, $e^{i(m_1\bk_1+m_2\bk_2)\c \bx}$ is in $\chi_{\bk_1}$ if and only if $(m_1+m_2)\equiv 1 \pmod 3$
 and $e^{i(m_1\bk_1+m_2\bk_2)\c \bx}$ is in $\chi_{-\bk_1}$ if and only if $(m_1+m_2)\equiv 2 \pmod 3$. It follows that $e^{i(m_1\bk_1+m_2\bk_2)\c \bx}$ must be in one and only one of $\chi_s$, $\chi_{\bk_1}$ and $\chi_{-\bk_1}$. This completes the proof.
\qed

Obviously, multiplications by elements in $\chi_s$ will map $\chi_s$, $\chi_{\bk_1}$, and $\chi_{-\bk_1}$ into themselves. To some extent, this shows functions in $\chi_s$, or to be specific, our potential $V(\bx)$ possesses higher symmetry. Besides, the decomposition above has the following symmetry: 

\begin{proposition}\label{prop-P}
If $f(\bx)$ is in $\chi_{\bk_1}$, then $f(-\bx)$ is in $\chi_{-\bk_1}$, and vice versa.
\end{proposition}

Thus, the transformation $\mcp[f](\bx) = f(-\bx)$ takes $\chi_{\bk_1}$ and $\chi_{-\bk_1}$ exactly to each other. \par

Secondly, we introduce rotation  eigen-subspaces of $\chi_s$, $\chi_{\bk_1}$, and $\chi_{-\bk_1}$ according to Fourier analysis. 
It is easy to get

\begin{equation}\label{eqn-R-on-k}
    R \bk_1 = \bk_2,\q R \bk_2 = -\bk_1-\bk_2,\q R(-\bk_1-\bk_2) = \bk_1,
\end{equation}
and
\begin{equation}\label{eqn-R-on-q}
     R \bq_1 = \bq_2,\q R \bq_2 = -\bq_1-\bq_2,\q R(-\bq_1-\bq_2) = \bq_1.
\end{equation}

Note the fact that:

\begin{lemma}\label{lem-L^0,+,--invariant-under-R}
$\chi_s,\chi_{\bk_1},\chi_{-\bk_1}$ are invariant function spaces of $\mcr$.
\end{lemma}

\Proof From (\ref{eqn-R-on-q}), clearly $\mcr$ maps $\chi_s$ to itself. Since 
$$\mcr(e^{i\bk_1\c \bx})=e^{i\bk_1\c R^*\bx}=e^{iR\bk_1\c \bx}=e^{i\bk_2\c \bx} = e^{i\bk_1\c \bx}e^{-i\bq_1\c \bx} $$ is in $\chi_{\bk_1}$, $\mcr$ maps $\chi_{\bk_1}$ to itself, too. It is similar for $\chi_{-\bk_1}$ with 
$$\mcr(e^{-i\bk_1\c \bx})=e^{-i\bk_2\c \bx} = e^{-i\bk_1\c \bx}e^{i\bq_1\c \bx} . $$
\qed

The next proposition gives detailed properties of these spaces with respect to the transformation $\mcr$ by separating $\chi_s$, $\chi_{\bk_1}$, and $\chi_{-\bk_1}$ into eigenspaces of $\mcr$. 
This decomposition helps to associate the rotation symmetry with conjugation symmetry, as in Proposition \ref{prop-C}.

\begin{proposition}\label{prop-R-decomposition}
The $\mcr$-invariant spaces $\chi_s$, $\chi_{\bk_1}$ and $\chi_{-\bk_1}$ have the following decomposition:
\begin{equation}\label{eqn-decomposition-for-L^0}
    \chi_s = \chi_{s,1} \oplus \chi_{s,\tau} \oplus \chi_{s,\overline{\tau}}
\end{equation}
\begin{equation}\label{eqn-decomposition-for-L^+}
    \chi_{\bk_1}= \chi_{\bk_1,1} \oplus \chi_{\bk_1,\tau} \oplus \chi_{\bk_1,\overline{\tau}}
\end{equation}
\begin{equation}\label{eqn-decomposition-for-L^-}
    \chi_{-\bk_1} = \chi_{-\bk_1,1} \oplus \chi_{-\bk_1,\tau} \oplus \chi_{-\bk_1,\overline{\tau}}
\end{equation}
where $\tau = e^{2\pi i/3}$ and $\chi_{\sigma,\xi} = \{ f \in \chi_{\sigma},\mcr[f](\bx) = \xi f(\bx) \}$, for $\sigma =s, \bk_1, -\bk_1,$ and $ \xi=1, \tau,\overline{\tau} $.
\end{proposition}

\Proof Take (\ref{eqn-decomposition-for-L^+}) as an example, others are similar.

According to (\ref{eqn-fourier-L^+}), let $(n_1,n_2)$ denote $$e^{i(\bk_1+n_1\bq_1+n_2\bq_2)\c \bx}=e^{i((n_1+n_2+1)\bk_1 + (2n_2-n_1)\bk_2)\c \bx}$$, then
\begin{equation}\label{eqn-R-on-L^+}
    (n_1,n_2) \stackrel{\mcr}{\longrightarrow} (-n_2-1,n_1-n_2)\stackrel{\mcr}{\longrightarrow} (n_2-n_1-1, -n_1-1)
\end{equation}
are all in $\chi_{\bk_1}$. Note that
$$(n_1,n_2) = \mcr(n_1,n_2) = (-n_2
-1,n_1-n_2),$$$$ (n_1,n_2) = \mcr^2(n_1,n_2) = (n_2-n_1-1, -n_1-1)$$ have no integer solutions. Thus, we can define the equivalence relation: $(n_1,n_2) \sim (-n_2-1,n_1-n_2) \sim (n_2-n_1-1, -n_1-1)$ and equivalence class $ S_+ = \Z^2/\sim$. 

Now it is clear that:

\begin{equation}\label{eqn-fourier-L^+_1}
    \begin{aligned}
    {
     \chi_{\bk_1,1} = \{f(\bx) \in \chi : f(\bx)=\sum_{(n_1,n_2)\in S_+}c(n_1,n_2)(e^{i(\bk_1+n_1\bq_1+n_2\bq_2)\c\bx} + } \\ {e^{iR(\bk_1+n_1\bq_1+n_2\bq_2)\c  \bx}  
      + e^{iR^2(\bk_1+n_1\bq_1+n_2\bq_2)\c \bx} ),\q c(n_1,n_2) \in \C \} }
    \end{aligned}
    \end{equation}
    
    \begin{equation}\label{eqn-fourier-L^+_tau}
   {
    \begin{aligned}
     \chi_{\bk_1,\tau} = \{f(\bx)\in \chi : f(\bx)=\sum_{(n_1,n_2)\in S_+}c(n_1,n_2)(e^{i(\bk_1+n_1\bq_1+n_2\bq_2)\c\bx} + \\ \overline{\tau}e^{iR(\bk_1+n_1\bq_1+n_2\bq_2)\c \bx}  
      + \tau e^{iR^2(\bk_1+n_1\bq_1+n_2\bq_2)\c \bx} ), \q c(n_1,n_2) \in \C\}
    \end{aligned} }
    \end{equation}

    \begin{equation}\label{eqn-fourier-L^+-bar-tau}
    {
    \begin{aligned}
     \chi_{\bk_1,\overline{\tau}} = \{f(\bx)\in \chi: f(\bx)=\sum_{(n_1,n_2)\in S_+}c(n_1,n_2)(e^{i(\bk_1+n_1\bq_1+n_2\bq_2)\c\bx} + \\ \tau e^{iR(\bk_1+n_1\bq_1+n_2\bq_2)\c \bx}  
      + \overline{\tau}e^{iR^2(\bk_1+n_1\bq_1+n_2\bq_2)\c \bx} ),\q c(n_1,n_2) \in \C\}
    \end{aligned} }
    \end{equation}  
    \qed
    
Again, given below is some information about symmetry properties between these subspaces.

\begin{proposition}\label{prop-C}
If $f(\bx)$ is in $\chi_{\pm\bk_1,\tau}$, then $\mcp\mcc[f](\bx)=\overline{f(-\bx)}$ is in $\chi_{\pm\bk_1,\overline{\tau}}$, and vice versa.
\end{proposition}

\section{Double Dirac cone in the band structure}\label{Double-cone}
In this section, we shall state the main theorem of the fourfold degeneracy and doubly conical structures at the $\Gamma$ point for the two-dimensional non-relativistic Sch\"odinger operator $H_V$ with super honeycomb lattice potentials. And a rigorous proof follows. Our proof is mainly inspired by the pioneering works \cite{RN18,RN84,RN77,RN17}. The proof is divided into two parts. First, we show that the fourfold degeneracy and a particular non-degenerate condition about eigenfunction are sufficient for the existence of double Dirac cones. Due to the higher degeneracy, we need deal with a more complicated bifurcation matrix. However, taking advantage of the higher symmetries and corresponding novel decomposition of working space in last section, we find that the bifurcation matrix can be pretty concise. Then we establish the existence of the degenerate point and the prescribed condition to complete the proof. Specifically, we first justify the prescribed conditions for shallow potentials, and then extend the justification for generic potentials.

\subsection{Main theorem of double Dirac cone}\label{main-thm}

The main theorem of our paper is as follows.

\begin{theorem}\label{thm-main}
    Let $V(\bx)$ be a super honeycomb lattice potential in the sense of Definition \ref{def-superHoneycomb}. 
    $H^{(\epsilon)} = -\Delta + \epsilon V(\bx)$ is the corresponding Schr\"odinger operator. 
    Then the following is true for all $\epsilon \in \R \setminus A $, where A is a discrete subset of $\R$:
    \begin{enumerate}
        \item there exists a real number $\mu_{_D}$ such that $\mu_D$ is an eigenvalue of $H^{(\epsilon)}$ on $\chi$ of multiplicity four. Namely,  there exists $b \in \N $ such that
        \begin{equation}\label{eqn-fourf}
            \mu_b(\bm{0}) < \mu_{_D}= \mu_{b+1}(\bm{0})  = \mu_{b+2}(\bm{0})  = \mu_{b+3}(\bm{0})
            = \mu_{b+4}(\bm{0})  < \mu_{b+5}(\bm{0}),
        \end{equation}
        \item there exists a number $v_{F}> 0$, such that for $\bk$ sufficiently small, the four spectral bands are of the form:
          \begin{equation}\label{eqn-double-cone}
            \begin{aligned}
              \mu_{b+1}( \bk) = \mu_{_D} - v_F| \bk |(1+\eta_1( \bk)) \\
              \mu_{b+2}( \bk) = \mu_{_D} - v_F| \bk |(1+\eta_2( \bk)) \\
              \mu_{b+3}( \bk) = \mu_{_D} + v_F| \bk |(1+\eta_3( \bk)) \\
              \mu_{b+4}( \bk) = \mu_{_D} + v_F| \bk |(1+\eta_4( \bk)) \\
            \end{aligned}
          \end{equation}where $\eta_j( \bk) = O(\v \bk \v)$, $j=1,2,3,4$. 
    \end{enumerate}
\end{theorem}

(\ref{eqn-double-cone}) is the strict description of the double Dirac cone mathematically. 
Thus, this theorem tells that there always exist two tangent cones with the same apex for the non-degenerate super honeycomb lattice potentials.

The rest of this section is proof of this theorem.
First, we give the conclusion that the fourfold degeneracy under some conditions always yields the double Dirac cone for super honeycomb lattice potentials.
Thus, our attention should be paid mainly to the existence of this condition and fourfold degeneracy.

\subsection{Fourfold degeneracy and double Dirac cone}\label{double-cones-after-degeneracy}

Due to (\ref{lem-symmetry}), if $\phi(\bx) \in ker(H_V - \mu I)$, then $\mcp\phi(\bx),\mcc\phi(\bx),\mcp \mcc \phi(\bx) \in ker(H_V - \mu I)$. 
That is to say, if $H_V$ does have an eigenvalue on $\chi_{\bk_1,\tau}$ of multiplicity one, then the fourfold degeneracy of $H_V$ on $\chi$ can be realized by symmetries. In this subsection, we want to verify that energy bands intersect conically in a small vicinity of this kind of eigenvalues under some necessary assumptions.

\begin{proposition}\label{thm-0-dirac-cone}
Let $V(\bx)$ be a super honeycomb lattice potential. Assume that:
\begin{enumerate}
    \item there exists a real number $\mu_{_D}$ and $b \in \N $ such that (\ref{eqn-fourf}) holds for energy bands of $H_V$,
    \item $\mu_{_D}$ is a simple eigenvalue on $\chi_{\bk_1,\tau}$ of $H_V$ with eigenfunction $\phi_1(\bx)$ and $$<\phi_1(\bx), \phi_1(\bx)>=1,$$
    \item $\bm{v}_{\sharp} =< \phi_1(\bx), \nabla \mcp \mcc [\phi_1](\bx) > \neq \bf{0} $.
\end{enumerate}
Then (\ref{eqn-double-cone}) holds for energy bands of $H_V$.
\end{proposition}

\begin{remark}
    The assumption that $\mu_{_D}$ is of multiplicity four is necessary, because for higher degeneracy, obviously more branches are included, and thus the bands near the $\Gamma$ point will be more complex.
    Hence, we need to verify this condition in the later subsections.
\end{remark}

\Proof
Using Proposition \ref{prop-P}, \ref{prop-C} and \ref{lem-symmetry}, we easily deduce that $\mu_{_D}$ is also a simple eigenvalue on $\chi_{\bk_1,\overline{\tau}}$, $\chi_{-\bk_1,{\tau}}$, $\chi_{-\bk_1,\overline{\tau}}$ with eigenfunctions $\phi_2(\bx)$, $\phi_3(\bx)$, and $\phi_4(\bx)$ :
\begin{equation}\label{eqn-phi2-4}
          \phi_2(\bx) = \mcp\mcc\phi_1(\bx),\q \phi_3(\bx)=\mcp\phi_1(\bx),\q \phi_4(\bx)=\mcc\phi_1(\bx), 
\end{equation}

Based on this, let us observe how the dispersion surfaces develop. First rewrite down the eigenvalue problem near $\bk = \bf{0}$. Assume $\bk$ is sufficiently small, the $\bk$ quasi-momentum eigenproblem (\ref{eqn-Hvk-eigen-problem-1})-(\ref{eqn-Hvk-eigen-problem-2}) is:
\begin{equation}\label{eqn-delta-k-0}
    H_V(\bk) \phi(\bx) = \mu(\bk) \phi(\bx), \q x\in \Omega
\end{equation}
\begin{equation}\label{eqn-delta-k-1}
    \phi(\bx + \buu)  =   \phi(\bx) ,\q \buu \in \bf{U}.
\end{equation}

Now let $\mu(\bk) = \mu_{_D} + \lambda $ and 
\begin{equation}\label{eqn-phi-k-x}
    \phi(\bx) = \alpha^j \phi_j(\bx) + \psi(\bx)
\end{equation}
with $\phi(\bx)\in \chi$ and $ \psi(\bx) \perp ker(H_V-\mu_{_D} I)$. 
We use the same superscript and subscript to represent summation over this script throughout the article. 
Since $ker(H_V-\mu_{_D} I)$ is a closed subspace of $\chi$, there exists projection operator $\mcq_{\parallel}$ from $\chi$ to $ker(H_V -\mu_{_D} I)$ and $\mcq_{\perp}$ from $\chi$ to $ker(H_V -\mu_{_D} I)^{\perp}$. 
It is trivial that $(H_V- \mu_{_D} I )\psi(\bx) \in ker(H_V- \mu_{_D} I)^{\perp}$.

Substitute (\ref{eqn-phi-k-x}) into (\ref{eqn-delta-k-0}):
\begin{equation}\label{eqn-before-project}
    (H_V - \mu_{_D} I ) \psi(\bx) = (2i\bk\cdot \nabla - |\bk|^2 +\lambda)( \alpha^j \phi_j(\bx) + \psi(\bx)).
\end{equation}

Project (\ref{eqn-before-project}) by $\mcq_{\parallel}$, 
\begin{equation}\label{eqn-parallel}
    0 = \mcq_{\parallel} (2i\bk \cdot \nabla - |\bk|^2 +\lambda)\alpha^j \phi_j(\bx) +  \mcq_{\parallel} 2i\bk \cdot \nabla \psi(\bx);
\end{equation}
by $\mcq_{\perp}$,
\begin{equation}\label{eqn-perp}
    (H_V - \mu_{_D} I ) \psi(\bx) = \mcq_{\perp}2i\bk \cdot \nabla\alpha^j \phi_j(\bx) + \mcq_{\perp}(2i\bk \cdot \nabla - |\bk|^2 +\lambda)\psi(\bx).
\end{equation}

Eigenvalue problem (\ref{eqn-delta-k-0})-(\ref{eqn-delta-k-1}) is equivalent to 
(\ref{eqn-parallel})-(\ref{eqn-perp}). First, solve $\psi(\bx)$ from (\ref{eqn-perp}, 
and then go back to (\ref{eqn-parallel}) to obtain $\lambda$ by linear approximation to complete the proof. 
This is exactly a Lyapunov-Schmidt reduction strategy.

Consider (\ref{eqn-perp}). Note that the resolvent operator $(H_V-\mu_{_D} I)^{-1}$ is a bounded map from $ker(H_V-\mu_{_D} I)^{\perp}$ to $\mcq_{\perp}(H^1_{per})$. Accordingly, 
\begin{equation*}
    \big{(}I - (H_V-\mu_{_D} I)^{-1} \mcq _{\perp}(2i\bk \cdot \nabla-|\bk|^2 + \lambda) \big{)} \psi(\bx) =(H_V-\mu_{_D} I)^{-1}\mcq_{\perp}  2ik \cdot  \nabla  \alpha^j   \phi_j(\bx) 
\end{equation*}
Let $\mca =  (H_V-\mu_{_D} I)^{-1} \mcq _{\perp}(2i\bk \cdot \nabla-|\bk|^2 + \lambda)$. With $|\bk|+ |\lambda|$ sufficiently small, operator norm of $\mca$ should be less than 1, which means $I-\mca$ is invertible, and $(I-\mca)^{-1}$ is bounded and we have
\begin{equation}\label{eqn-psi}
    \psi(\bx) = (I - \mca)^{-1} (H_V-\mu_{_D} I)^{-1}\mcq_{\perp} 2i\bk \cdot \nabla \alpha^j\phi_j(\bx).
\end{equation}

Let $\mct_j[\bk,\lambda](\bx) = (I - \mca)^{-1} (H_V-\mu_{_D} I)^{-1}\mcq_{\perp} 2i\bk \cdot \nabla \phi_j(\bx)$. It is a bounded map from a sufficient small neighborhood of $(\bk, \lambda)=(\bf{0},0)$ in $\R^2 \times \C$ to $H^1_{per}$. Its norm is less than $C(|\bk| + |\lambda|)$ in the small neighborhood. Thus,  $\parallel \nabla \mct_j[\bk,\lambda](\bx)) \parallel_{L^2} \leq C(|\bk| + |\lambda|) $. Rewrite (\ref{eqn-psi}) as 
\begin{equation}\label{eqn-psi---}
    \psi(\bx) = \alpha^j\mct_j[\bk,\lambda](\bx)
\end{equation}
Substituting into (\ref{eqn-parallel}),
\begin{equation}
\mcq_{\parallel} \bigg{(}(2i\bk \cdot \nabla - |\bk|^2 +\lambda) \phi_j(\bx)  +   2i\bk \cdot \nabla \mct_j[\bk,\lambda](\bx)\bigg{)}\alpha^j = 0
\end{equation}
This equation has a nonzero solution $\{\alpha^j\}$ when the matrix
\begin{equation}\label{eqn-M}
     M(\bk, \lambda) = (<\phi_l(\bx), (2i\bk \cdot \nabla - |\bk|^2 +\lambda) \phi_j(\bx) + 2i\bk \cdot \nabla \mct_j[\bk,\lambda](\bx) > )_{l,j}
\end{equation}
has a nonzero solution, which is equivalent to
\begin{equation}\label{eqn-det}
   det(M(\bk,\lambda))=0
\end{equation} 

Our aim is to solve $\lambda$ from this equation. Divide $M(\bk,\lambda)$ into two parts $M_0$ and $M_1$:
\begin{equation}\label{eqn-M_0}
    M_0(\bk,\lambda) = (< \phi_l(\bx) , ( \lambda +2i\bk\cdot\nabla)\phi_j(\bx)>)_{l,j}
\end{equation}
\begin{equation}\label{eqn-M_1}
    M_1 (\bk,\lambda)= (- |\bk|^2 <\phi_l(\bx),  \phi_j(\bx) > )_{l,j}  + (<\phi_l(\bx), 2i\bk \cdot \nabla  \mct_j[\bk,\lambda](\bx) > )_{l,j}.
\end{equation}

Fix $\bk$, let $(\bk,\lambda(\bk))$ be the solution of (\ref{eqn-det}). Taking advantage of (\ref{lem-mu-lipschitz}), $\lambda(\bk)$ should be Lipschitz continuous with $\bk$ on each branch of the solutions. Thus, $M_1$ is of order $O(|\bk|^2)$. That is to say, $M_1$ will only contribute a term of order $(|\bk|^8)$ to $det(M(\bk,\lambda))$ when $det(M(\bk,\lambda)) = 0$. We first solve the truncated equation of (\ref{eqn-det}):
\begin{equation}
    det(M_0(\bk,\lambda)) = 0
\end{equation}
The $M_0$ here, is the linear truncation of the original question, and is called a bifurcation matrix. 

{We already have $\langle \phi_1(\bx), \phi_1(\bx) \rangle = 1$. Because that the subspaces are orthogonal, $\langle \phi_l(\bx), \phi_j(\bx)\rangle = 0$ for $l\neq j$. And it is also obvious that $\langle \phi_l(\bx), \phi_l(\bx) \rangle = 1$ for all $l$ due to (\ref{eqn-phi2-4}) and some simple calculations.} Note that 
\begin{equation}\label{eqn-elment-in-birfucation}
    \begin{aligned}
        <\phi_1(\bx), 2i\bk\cdot \nabla \phi_2(\bx)> &= 2i \bk \cdot \bm{v}_{\sharp} \\
        <\phi_2(\bx), 2i\bk\cdot \nabla \phi_1(\bx)> &= \overline{<\phi_1(\bx), 2i\bk\cdot \nabla \phi_2(\bx)>} =\overline{2i \bk \cdot \bm{v}_{\sharp}}, \\
        <\phi_3(\bx), 2i\bk\cdot \nabla \phi_4(\bx)> &= <\phi_1(\bx), - 2i\bk\cdot \nabla \phi_2(\bx)> = - 2i \bk \cdot \bm{v}_{\sharp}, \\
        <\phi_4(\bx), 2i\bk\cdot \nabla \phi_3(\bx)> &= <\phi_2(\bx), - 2i\bk\cdot \nabla \phi_1(\bx)> = - \overline{2i \bk \cdot \bm{v}_{\sharp}}.
    \end{aligned}
\end{equation}

Besides, we also have the following results.
\begin{proposition}
\label{lem-item-in-M-1}
    For $\psi_j(\bx)$ satisfies $\mcr\psi_j(\bx) = \xi_j \psi_j(\bx)$, $\xi_j\in\{1,\tau,\overline{\tau}\}$, $j=1,2$ $$    \xi_1 = \xi_2  \q \Rightarrow  \q <\psi_1(\bx),  \nabla \psi_2(\bx)> = \bf{0} $$
    
\end{proposition}
    
    \Proof
    Since the transformation $\mcr$ is unit, if $\xi_1=\xi_2$, we can attain that
    \begin{equation*}
        \begin{aligned}
         <\psi_1(\bx), \nabla\psi_2(\bx)> &= <\mcr \psi_1(\bx), \mcr \nabla \psi_2(\bx)> = \overline{\xi_1} <\psi_1(\bx) , \nabla_{R^*\bx} \psi_2(R^*\bx)> \\
         & = \overline{\xi_1} < \psi_1(\bx), R^*\nabla \mcr\psi_2(\bx) > = \overline{\xi_1} \xi_2 < \psi_1(\bx), R^*\nabla \psi_2(\bx) > \\
         & = \overline{\xi_1} \xi_2  R^* < \psi_1(\bx), \nabla \psi_2(\bx) > = R^* < \psi_1(\bx), \nabla \psi_2(\bx) >
        \end{aligned}
    \end{equation*}
    Here $R^*$ is the rotation matrix. Because $1$ is not an eigenvalue of $R^*$,   $$<\psi_1(\bx),  \nabla \psi_2(\bx)> = \bf{0}.$$ Besides, this equation tells us that $<\psi_1(\bx),  \nabla \psi_2(\bx)> $ must be $\bf{0}$ or an eigenvector of $R^*$.
    \qed
    
    From the proof of this lemma, we know $\bm{v}_{\sharp}$ is an eigenvector of $R^*$ with eigenvalue $\tau$. Thus it can be written as : 
    \begin{equation}\label{eqn-v_F}
        \bm{v}_{\sharp} =  \frac{v_F}{2} \begin{pmatrix} 1 \\ i\end{pmatrix} e^{i\theta}.
    \end{equation}
    We can choose an appropriate $\theta$ such that $v_F > 0$  because $\bm{v}_{\sharp} \neq \bf{0}$.

    \begin{proposition}
    
    \label{lem-item-in-M-1-2}
    For $\phi(\bx) \in \chi_{\pm\bk_1}$, each component of $\nabla\phi(\bx)$ is in $\chi_{\pm\bk_1}$ too.
    \end{proposition}
    \Proof
    If $\phi_(\bx) \in \chi_{\pm\bk_1}$, it can be written as $\phi_(\bx)= e^{\pm i k_1\cdot\bx}p(\bx)$, where $p(\bx) \in \chi_s$. Therefore each component of $\nabla \phi(\bx) = \pm i \bk_1 e^{\pm ik_1 \cdot \bx} p(\bx) + e^{\pm ik_1 \cdot \bx} \nabla p(\bx)$
    is in $\chi_{\pm\bk_1}$ too.
    \qed
        
Now using (\ref{lem-item-in-M-1}), (\ref{lem-item-in-M-1-2}), and again the orthogonality of subspaces, 
we can obtain the bifurcation matrix

\begin{equation}
    M_0(\bk,\lambda) = \begin{pmatrix}
     \lambda & 2i\bk\cdot \bm{v}_{\sharp} & 0 & 0 \\
       \overline{2i\bk\cdot \bm{v}_{\sharp}}& \lambda & 0 & 0 \\
        0 & 0 & \lambda & -2i\bk\cdot \bm{v}_{\sharp} \\
         0 & 0 & -\overline{2i\bk\cdot \bm{v}_{\sharp}}  & \lambda   \\
    \end{pmatrix}.
\end{equation}
And $det(M_0(\bk,\lambda)) = (\lambda^2 - 4|\bk \cdot \bm{v}_{\sharp}|^2)^2 $. Now solve the whole equation (\ref{eqn-det}).  Thanks to all the discussion above, it can be written in the form that:
\begin{equation}
    (\lambda^2 - 4|\bk \cdot \bm{v}_{\sharp}|^2)^2 + O(|\bk|^8) = 0
\end{equation}
This equation's solution $(\bk, \lambda(\bk))$ gives four branches of the dispersion surfaces by $\mu(\bk) = \mu + \lambda(\bk)$. Due to (\ref{eqn-v_F}), for $\bk \in \R^2$, $2 |\bk \cdot \bm{v}_{\sharp}| = v_F|\bk|$.
Thus, for $|\bk|$ sufficiently small, they are exactly:
\begin{equation}
  \begin{aligned}
   \mu_{b+1}(\bk) = \mu_{_D} - v_F|\bk|(1+\eta_1( \bk)) \\
    \mu_{b+2}(\bk) = \mu_{_D} - v_F|\bk|(1+\eta_2( \bk)) \\
    \mu_{b+3}(\bk) = \mu_{_D} + v_F|\bk|(1+\eta_3( \bk))\\
    \mu_{b+4}(\bk) = \mu_{_D} + v_F|\bk|(1+\eta_4( \bk)) 
  \end{aligned}
\end{equation}
where $\eta_j(\bk) = O(\v \bk \v)$, $j=1,2,3,4$. 
\qed

\subsection{Fourfold degeneracy with shallow potentials}

Due to the discussion in section 2.1 , the left thing to do is finding a single eigenvalue of $H^{(\epsilon)}$ on $\chi_{\bk_1,\tau}$ which is not an eigenvalue on $\chi_{\pm\bk_1,1}$ and $\chi_s$ for all $\epsilon$ except a discrete set. In this subsection, we discuss the situation of $\epsilon$ sufficiently small. 

First, take $\epsilon=0$. The eigenvalue problem is
\begin{equation}
    -\Delta \phi(\bx) = \mu \phi(\bx)  ,\q \bx \in \R^2,
\end{equation}
\begin{equation}
    \phi(\bx + \buu) = \phi(\bx) ,\q \buu \in \bf{U}.
\end{equation}

The operator $-\Delta$ is positive semi-definite on $H^1_{per}$. It is quite easy to know its spectrum from Sturm-Liouville Theorem and Fourier series' presentations. Pay attention to the first eight eigenvalues:
$$\mu_1 = 0 < \mu_2 = \mu_3 = \mu_4 = \mu_5 = \mu_6 = \mu_7 = |\bk_1|^2 < \mu_8 $$
Obviously, a group of eigenfunctions for $\mu_2$-$\mu_7$ are:
\begin{equation*}
  \{ e^{i\bk_1\cdot\bx}, \q e^{iR\bk_1\cdot\bx} ,\q e^{iR^2\bk_1\cdot\bx} ,\q
   e^{-i\bk_1\cdot\bx}, \q  e^{-iR\bk_1\cdot\bx} , \q e^{-iR^2\bk_1\cdot\bx} \}
\end{equation*}
After some linear combinations, it is equivalent to:
\begin{equation}\label{eqn-episilon-0}
     \begin{aligned}
     \{\q & \phi_1^{(0)}(\bx)=\frac{1}{\sqrt{3}} ( e^{i\bk_1\cdot\bx}+ \overline{\tau} e^{iR\bk_1\cdot\bx} + \tau e^{iR^2\bk_1\cdot\bx})  & \in \q& \chi_{\bk_1,\tau},  
     \\ &\phi_2^{(0)}(\bx) = \frac{1}{\sqrt{3}} ( e^{i\bk_1\cdot\bx}+  \tau e^{iR\bk_1\cdot\bx} + \overline{\tau} e^{iR^2\bk_1\cdot\bx})  &\in \q & \chi_{\bk_1,\overline{\tau}}, \\
     & \phi_3^{(0)}(\bx) =\frac{1}{\sqrt{3}} (  e^{-i\bk_1\cdot\bx}+ \overline{\tau} e^{-iR\bk_1\cdot\bx} + \tau e^{-iR^2\bk_1\cdot\bx})  &\in\q &\chi_{-\bk_1,\tau}, \\
     & \phi_4^{(0)}(\bx) = \frac{1}{\sqrt{3}} ( e^{-i\bk_1\cdot\bx}+  \tau e^{-iR\bk_1\cdot\bx} + \overline{\tau} e^{-iR^2\bk_1\cdot\bx}) & \in \q &\chi_{-\bk_1,\overline{\tau}}, \\
     & \phi_5^{(0)}(\bx) = \frac{1}{\sqrt{3}} (
     e^{i\bk_1\cdot\bx}+ e^{iR\bk_1\cdot\bx} + e^{iR^2\bk_1\cdot\bx} ) &\in \q &\chi_{\bk_1,1},\\
     & \phi_6^{(0)}(\bx) = \frac{1}{\sqrt{3}} (
     e^{-i\bk_1\cdot\bx}+ e^{-iR\bk_1\cdot\bx} + e^{-iR^2\bk_1\cdot\bx})  &\in \q &\chi_{-\bk_1,1}, \q \}
    \end{aligned} .
\end{equation}

Secondly, for $\epsilon$ small enough, the $2^{nd}-7^{th}$ eigenvalues of $H^{(\epsilon)}$ on $\chi$ must also be eigenvalues on $\chi_{\bk_1}$ or $\chi_{-\bk_1}$ and separated from other eigenvalues by the continuity of the eigenvalues in $\epsilon$. Thus, whether the fourfold degeneracy on $\chi_{\pm\bk_1, \tau}$ and $\chi_{\pm\bk_1,\overline{\tau}}$ given by parity and conjugation symmetry and twofold degeneracy on $\chi_{\pm\bk_1, 1}$ given by conjugation symmetry will separate is most concerned about in this subsection.

\begin{proposition}\label{thm-0-episilon-small}
Assume that $V(\bx)$ is a super honeycomb lattice potential. And $H^{(\epsilon)} = -\Delta + \epsilon V(\bx)$ is the corresponding Schr\"odinger operator.
Then there exists an $\epsilon_0>0$ such that, for all $\epsilon\in (-\epsilon_0,\epsilon_0) \setminus \{0\}$, there exists  $\mu_{_D}^{(\epsilon)}$ and $\bm{v}_{\sharp}^{(\epsilon)} \neq \bf{0}$ satisfies:
\begin{enumerate}
    \item  $\mu_{_D}^{(\epsilon)}$ is an eigenvalue of $H^{(\epsilon)}$ on $\chi$ of multiplicity four, 
    \item $\mu_{_D}^{(\epsilon)}$ is a simple eigenvalue on $\chi_{\bk_1,\tau}$ with eigenfunction $\phi_1^{(\epsilon)}(\bx)$,
    \item $\bm{v}_{\sharp}^{(\epsilon)} =<\phi_1^{(\epsilon)}(\bx), \nabla \mcp  \mcc \phi_1^{(\epsilon)}(\bx) > .$ 
\end{enumerate}
\end{proposition}

\begin{remark}
    From Lemma \ref{lem-symmetry}, we know that the conclusion 1 and 2 can deduce that $\mu_{_D}^{(\epsilon)}$ is not an eigenvalue on $\chi_{\pm\bk_1, 1}$.
\end{remark}

\Proof
The eigenvalue problem with shallow potentials on $\chi$ is
\begin{equation}\label{eqn-episilon-eigen-problem}
    (-\Delta + \epsilon V(\bx)) \phi^{(\epsilon)}(\bx) = \mu^{(\epsilon)} \phi^{(\epsilon)}(\bx), \q \bx \in \R^2
\end{equation}
with $\phi^{(\epsilon)}(\bx) \in \chi$. Due to discussion above and Lemma \ref{lem-mu-lipschitz}, only need to concern about the second to seventh eigenvalues. 

Taking advantage of the discussion in section 2.1, we can limit this problem to each space $\chi_{\sigma, \xi}$, where $\sigma \in \{ \bk_1,-\bk_1\}$ and $\xi \in \{ 1,\tau, \overline{\tau}\}$. Now $\phi^{(\epsilon)}(\bx) \in \chi_{\sigma, \xi}$. Analogous to the process in the proof of the last theorem, we rewrite the eigenvalue problem and divide it into two parts by projections. 

Let $\mu^{(\epsilon)} = \mu^0 + \lambda$, $\phi^{(\epsilon)}(\bx) = \phi^{(0)}(\bx) + \psi(\bx)$. Here $\mu^0 = |\bk_1|^2$ and $\phi^{(0)}(\bx)$ is the corresponding eigenfunction on $\chi_{\sigma, \xi}$ in  (\ref{eqn-episilon-0}). $\psi(\bx) \in span\{\phi^{(0)}(\bx)\}^{\perp} = ker(H^{(0)} - \mu^0I)^{\perp}$. Introduce new projection operators $\mcq_{\parallel}$ from $\chi_{\sigma, \xi}$ to $ker(H^{(0)} - \mu^0I)$ and $\mcq_{\perp}$ from $\chi_{\sigma, \xi}$ to $ker(H^{(0)} - \mu^0I)^{\perp}$. Perform $\mcq_{\parallel}$ and $\mcq_{\perp}$ on (\ref{eqn-episilon-eigen-problem}) to obtain
\begin{equation}\label{eqn-episilon-paralle}
     \epsilon \mcq_{\parallel} V(\bx) \psi(\bx) = \mcq_{\parallel} (\lambda - \epsilon V(\bx)) \phi^{(0)}(\bx) ,
\end{equation}
and
\begin{equation}\label{eqn-episilon-perp}
    (H^{(0)} - \mu^0 I) \psi(\bx) = \epsilon \mcq_{\perp} V(\bx)\phi^{(0)}(\bx) - \mcq_{\perp}(\epsilon V(\bx) - \lambda) \psi(\bx).
\end{equation}
These two equations are equivalent to the original equation, so only need to solve them. Note that $(H^{(0)} - \mu^0 I)^{-1}$ is a bounded operator from $ker(H^{(0)} - \mu^0I)^{\perp}$ to $H^1_{per}$. Thus,
\begin{equation}
     \big{(}I + (H^{(0)} - \mu^0 I)^{-1}\mcq_{\perp}(\epsilon V(\bx) - \lambda) \big{)}\psi(\bx)
     = \epsilon (H^{(0)} - \mu^0 I)^{-1} \mcq_{\perp} V(\bx)\phi^{(0)}(\bx) 
\end{equation}
Assume $|\epsilon|+|\lambda|$ is small enough, $\big{(}I + (H^{(0)} - \mu^0 I)^{-1}\mcq_{\perp}(\epsilon V(\bx) - \lambda) \big{)}^{-1}$ exists and is bounded. Use notation
$$\mct[\epsilon ,\lambda](\bx) = \big{(}I + (H^{(0)} - \mu^0 I)^{-1}\mcq_{\perp}(\epsilon V(\bx) - \lambda) \big{)}^{-1} (H^{(0)} - \mu^0 I)^{-1} \mcq_{\perp} V(\bx) \phi^{(0)}(\bx).  $$

$\mct$ is bounded in $H^1_{per}$ when $|\epsilon| + |\lambda|$ is small enough. Substituting $\psi(\bx)=\epsilon \mct[\epsilon,\lambda](\bx)$ into (\ref{eqn-episilon-paralle}), we have
\begin{equation}
    \epsilon ^ 2 \mcq_{\parallel} V(\bx) \mct[\epsilon,\lambda](\bx) = \mcq_{\parallel} (\lambda - \epsilon V(\bx))\phi^{(0)}(\bx)
\end{equation}
It is the same with:
\begin{equation*}
    \epsilon ^ 2 <V(\bx) \mct[\epsilon,\lambda](\bx), \phi^0(\bx)> =<(\lambda - \epsilon V(\bx))\phi^{(0)}(\bx) ,\phi^{(0)}(\bx)>
\end{equation*}
Solve it to attain
\begin{equation}
    \lambda(\epsilon) = <V(\bx)\phi^{(0)}(\bx), \phi^{(0)}(\bx)> \epsilon  +  <V(\bx) \mct[\epsilon,\lambda](\bx), \phi^{(0)}(\bx)>\epsilon^2
\end{equation}
Thus, we solve out a $\lambda(\epsilon)$ of order $O(|\epsilon|)$. This means there exactly exists a Floquet-Bloch eigenpair $(\mu_{_D}{(\epsilon)}, \phi^{(\epsilon)}(\bx))$ with $\mu_{_D}^{(\epsilon)}$ changing in order $O(|\epsilon|)$ on $\chi_{\sigma, \xi}$.

First we take $\sigma=\bk_1$ and $\xi = \tau$, consider $\phi^{(\epsilon)}_1(\bx) = \phi^{(0)}_1(\bx) + \epsilon \mct[\epsilon,\lambda](\bx)$. Provided $|\epsilon|$ sufficiently small, $\parallel\mct[\epsilon,\lambda(\epsilon)]\parallel_{H^1_{per}}$ is of order $O(|\epsilon)|$, therefore
\begin{equation*}
    \bm{v}_{\sharp}^{(\epsilon)} = <\phi^{(0)}_1(\bx), \nabla \mcp\mcc \phi^{(0)}_1(\bx)> + O(|\epsilon|) \neq 0.
\end{equation*}

Secondly, traversing all $\sigma$ and $\xi$, the six eigenpairs on $\chi_{\sigma, \xi}$ give the second to seventh eigenvalues for the original problem on $\chi$ when $\epsilon$ is quite small. As mentioned above, four on $\chi_{\bk_1,\tau}$, $\chi_{\bk_1,\overline{\tau}}$, $\chi_{-\bk_1,\tau}$ and $\chi_{\bk_1,\overline{\tau}}$ are bound to each other due to symmetry. The same for $\chi_{\bk_1,1}$ and $\chi_{-\bk_1,1}$. After some basic calculations, the $\lambda(\epsilon)$ on $\chi_{\sigma, \xi}$ is
\begin{equation}\label{eqn-lamda-L-sigma-eta}
    \lambda_{\sigma,\xi}(\epsilon) = (c_1 + (\xi + \overline{\xi})c_2) \epsilon + O(|\epsilon|^2),
\end{equation}
where $c_1 = <1, V(\bx)>$ and $c_2= <e^{i\bq_1\cdot\bx}, V(\bx)> $. $c_2 \neq 0$ is the non-degeneracy condition of super honeycomb lattice potentials.
It is obvious that $\lambda_{+,\tau} = \lambda_{+,\overline{\tau}} = \lambda_{-,\tau} = \lambda_{-,\overline{\tau}} \neq\lambda_{+,1} = \lambda_{-,1}$. This explains these six branches will separate into fourfold and twofold for super honeycomb lattice potentials.
\qed

\begin{proposition}
With conditions in Proposition \ref{thm-0-episilon-small} and $\epsilon$ sufficiently small, for the eigenvalue problem (\ref{eqn-episilon-eigen-problem}) on $\chi$, if $<e^{i\bq_1\cdot\bx}, V(\bx)>$ is positive, then
$$\mu^{\epsilon}_2 = \mu^{\epsilon}_3 = \mu^{\epsilon}_4 = \mu^{\epsilon}_5 < \mu^{\epsilon}_6 = \mu^{\epsilon}_7, $$ 
and if $<e^{i\bq_1\cdot\bx}, V(\bx)> $ is negative, then $$\mu^{\epsilon}_2 = \mu^{\epsilon}_3 < \mu^{\epsilon}_4 = \mu^{\epsilon}_5 = \mu^{\epsilon}_6 = \mu^{\epsilon}_7 $$
\end{proposition}

\Proof
It is easy to verify using (\ref{eqn-lamda-L-sigma-eta}).
\qed

\subsection{Proof of the main theorem}

Last subsection gives the result of shallow potentials, and this subsection briefly shows the method to verify Theorem \ref{thm-main}, the generic case.   
The key strategy is constructing an analytic function $\mce(\mu, \epsilon)$ on each $\chi_{\sigma, \xi}$, whose zero points are eigenvalues of $H_V$ on function spaces we concern. 
This function is actually the determinant of infinite dimensional linear operator $H^{(\epsilon)} = -\Delta + \epsilon V(\bx)$ by some renormalization using trace class.
Because of the symmetry in Lemma \ref{lem-symmetry} again, obviously the eigenvalue should exist simultaneously on $\chi_{\bk_1,\tau}$, $\chi_{\bk_1,\overline{\tau}}$, $\chi_{-\bk_1,\tau}$ and $\chi_{-\bk_1,\overline{\tau}}$. 
And this eigenvalue should be different from those on other subspaces.
The main work is focused on establishing this analytic function and prove the three conditions in Proposition \ref{thm-0-dirac-cone}.

Without loss of generality, let us assume that $0 \leq V(\bx) \leq V_m$. On each $\chi_{\sigma, \xi}$, the eigenvalue problem is:
$$
    (-\Delta + \epsilon  V(\bx)  ) \Phi(\bx) = \mu \Phi(\bx).
$$ In this subsection, we consider $\epsilon\in \C$. First consider $\epsilon$ with nonnegative real part $Re(\epsilon) \geq 0$. Our aim is to derive an operator whose determinant can be defined. Hence, we rewrite it as
\begin{equation}\label{eqn-rewrite-1}
    (I - \Delta + \epsilon V(\bx)) \Phi(\bx) = (\mu +1) \Phi(\bx),
\end{equation}
Note that $(I - \Delta + \epsilon V(\bx))$ is invertible. Let $\mca(\epsilon)= (I - \Delta + \epsilon V(\bx))^{-1}$, then 
\begin{equation}\label{eqn-rewrite-2}
    (I - (\mu+1)\mca(\epsilon)) \Phi(\bx) = 0.
\end{equation}

The $j^{th}$ eigenvalue $\lambda_j$ on $\chi $ of $\mca(\epsilon)$ is asymptotic to $j^{-1}$. Thus, it is a Hilbert-Schmidt operator:
$$\parallel \mca^2(\epsilon) \parallel = \sum_j | \lambda_j |^{-2} \sim \sum_j |j|^{-2} < \infty .$$

Now a determinant for $(I-(\mu+1)\mca (\epsilon))$ can be defined through a regularized way:
\begin{equation}
    det_2(I - (\mu+1)\mca(\epsilon)) = det (I + R_2((\mu+1)\mca(\epsilon))).
\end{equation}
The right-hand side is Fredholm determinant. It is well-defined because the regularized form:
\begin{equation}
    R_2((\mu+1)\mca(\epsilon)) = (I+(\mu+1)\mca(\epsilon))e^{-(\mu+1)\mca(\epsilon)} - I
\end{equation}
is a trace class. 

We already have the following lemma.

\begin{lemma}
    For all $\sigma \in \{s,\bk_1,-\bk_1\}$ and $\xi\in \{ 1, \tau, \overline{\tau}\}$, the following is true:
    \begin{enumerate}
        \item $\epsilon \to \mca(\epsilon)$ is an analytic mapping from $\{ \epsilon \in \C, Re(\epsilon) \geq 0 \}$ to the space of Hilbert-Schmidt operators on $\chi_{\sigma, \xi}$,
        \item The regularized determinant on $\chi_{\sigma, \xi}$ 
            $$ \mce_{\sigma, \xi}(\mu, \epsilon)= det_2(I - (\mu+1)\mca(\epsilon))
            $$
            is analytic for both $\mu$ and $\epsilon$ with $Re(\epsilon) \geq 0$,
        \item For $\epsilon$ real and nonnegative, $\mu$ is a $\chi_{\sigma, \xi}$ eigenvalue for $H_V$ of multiplicity $m$ if and only if it is a zero of $\mce_{\sigma, \xi}(\mu,\epsilon)$ of multiplicity $m$.
    \end{enumerate}

\end{lemma}

Therefore we first need to prove that except a discrete set in $\R$ for $\epsilon$, $\mce_{\bk_1,\tau}$ has a simple zero $(\epsilon, \mu)$ which is not a zero of $\mce_{\pm\bk_1,1}$ and $\mce_{s,\xi}$ for all $\xi$. Then prove that the eigenfunction on $\chi_{\bk_1, \tau}$ corresponding to this $\mu$ has nonzero $\bm{v}_{\sharp}$, which is defined in Proposition \ref{thm-0-dirac-cone}. This is totally the same with the proof in \cite{RN18}, based on the previous subsection's conclusion about shallow potentials. The only difference is about symmetry, which will not influence the proof.

For $\epsilon$ with real part negative, just replace (\ref{eqn-rewrite-1}) with
\begin{equation}
    (I - \Delta + \epsilon (V(\bx)-V_m) ) \Phi(\bx) = (\mu +1 - \epsilon V_m) \Phi(\bx)
\end{equation}
and replace (\ref{eqn-rewrite-2}) with 
\begin{equation}
    (I -(\mu +1 - \epsilon V_m) \Tilde{\mca}(\epsilon)\Phi(\bx) = 0,
\end{equation}
where $\Tilde{\mca}(\epsilon) = (I - \Delta + \epsilon (V(\bx)-V_m) )^{-1} $. The rest is similar.

\section{Instability under symmetry breaking perturbations}\label{Perturbations}

We already show the existence of double cones for the operator $H_V$ with a super honeycomb lattice potential $V(\bx)$. This section focuses on what will happen if the additional translation symmetry of the potential is broken. In other words, we investigate the behaviour of the band structures of $H_V$ after some perturbations.

\subsection{Perturbations  breaking additional translation symmetry }

\begin{figure}[htbp]
    \centering
    \subfigure[]{\includegraphics[width=6cm]{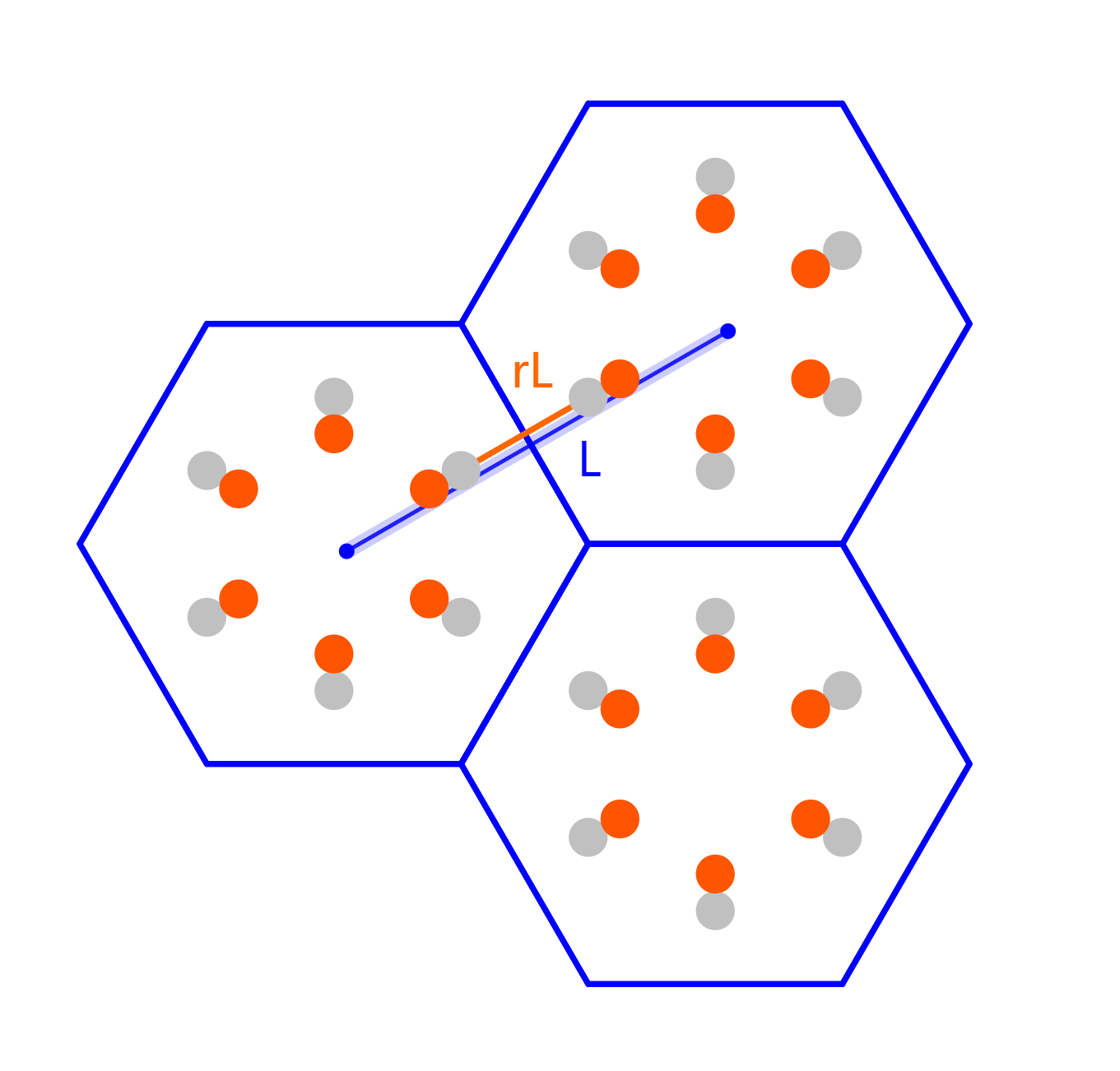}}
    \subfigure[]{\includegraphics[width = 6cm]{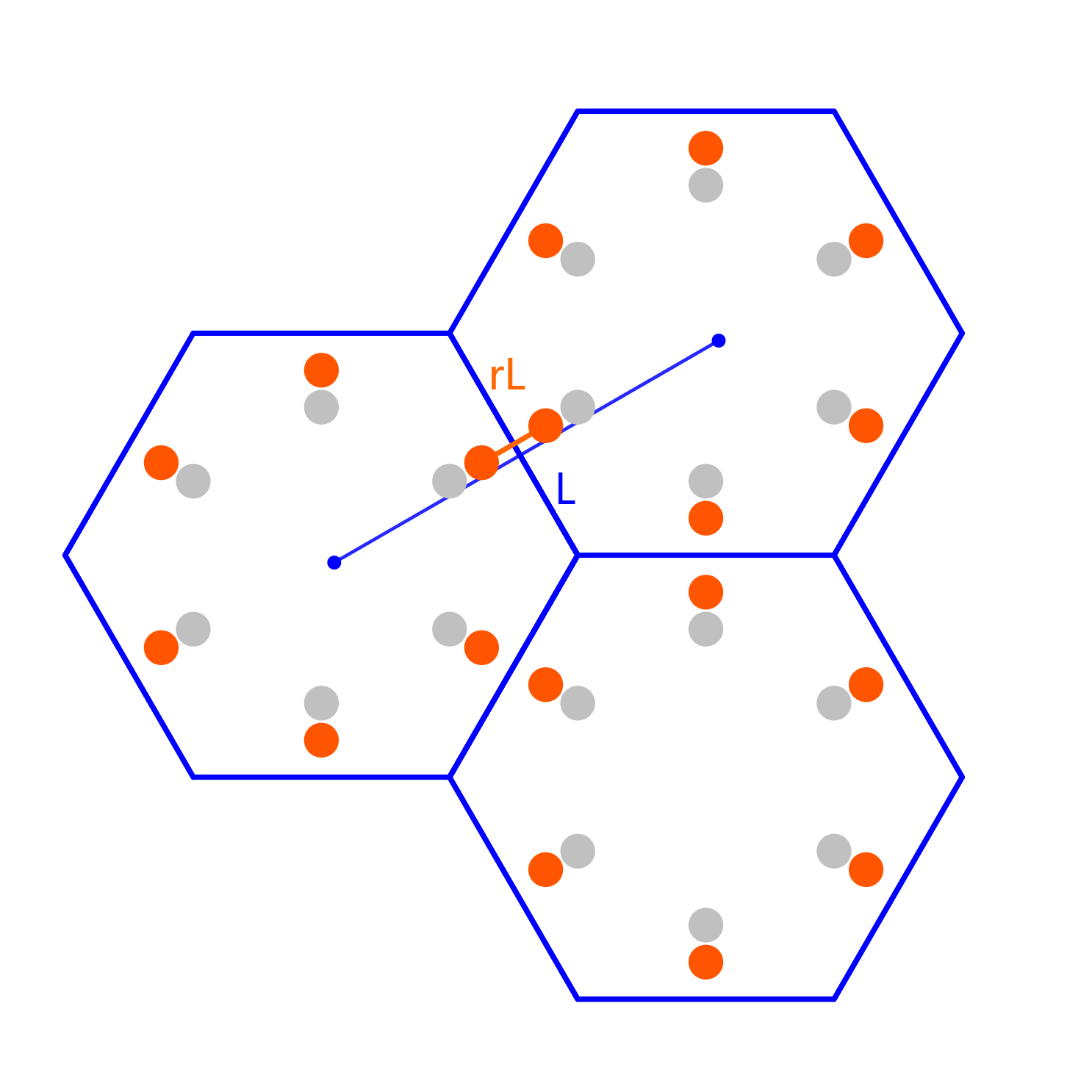}}
    \caption{Perturbed lattice based on a super honeycomb lattice: (a) the shrunk hexagonal lattice, and (b) the expanded hexagonal lattice. Grey points show the structure of super honeycomb lattice, and red 
    points show the perturbed case.
    The scalar $r$ is the ratio of the distance of the components in a dimer to the length of periodicity of honeycomb lattice. }
    \label{figure-perturbation}
\end{figure}

\begin{figure}
    \centering
    \subfigure[]{\includegraphics[width=4cm]{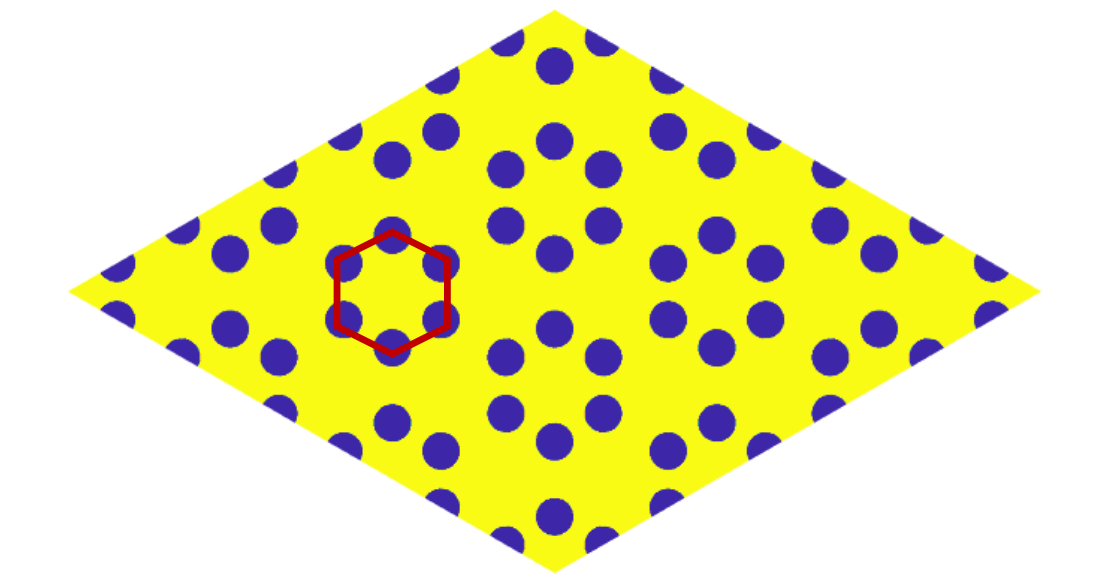}}
    \subfigure[]{\includegraphics[width=4cm]{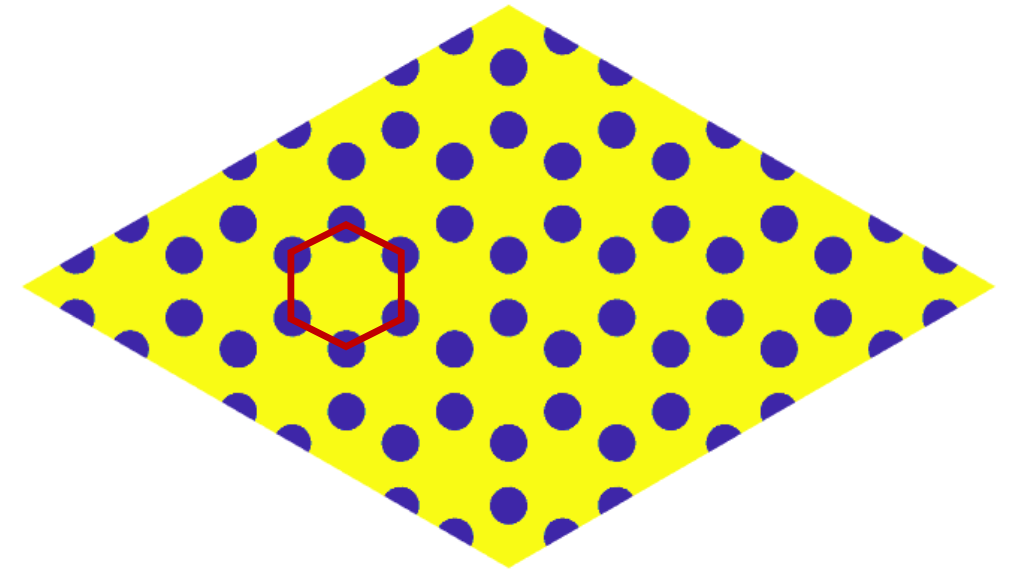}}
    \subfigure[]{\includegraphics[width=4cm]{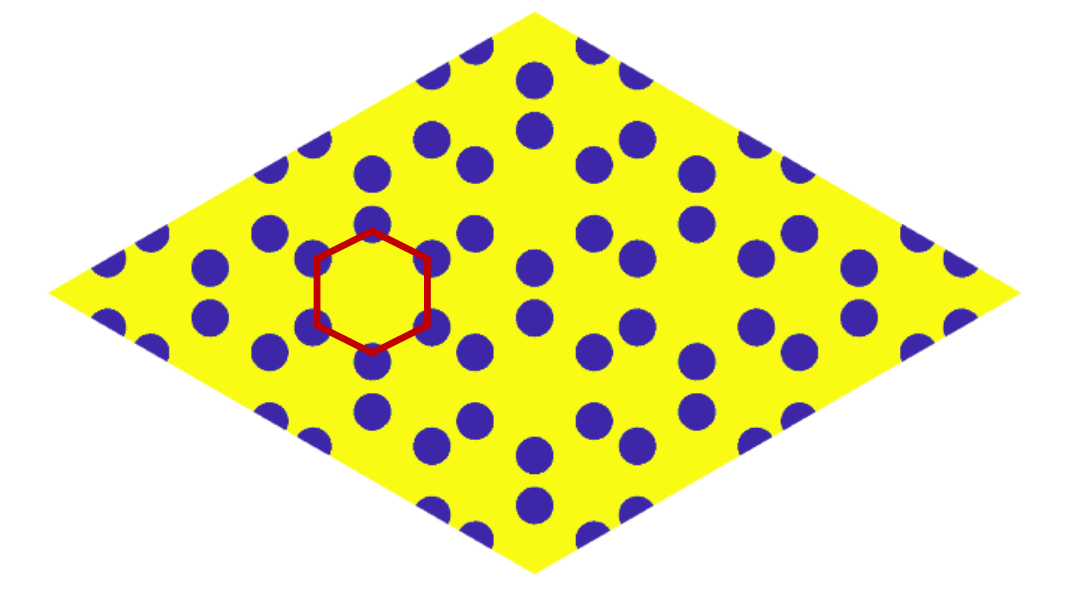}}\\
    \subfigure[]{\includegraphics[width=4cm]{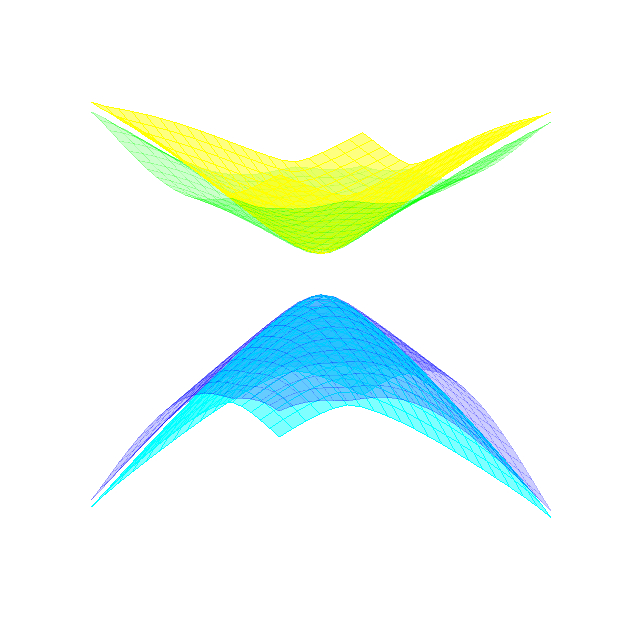}}
    \subfigure[]{\includegraphics[width=4cm]{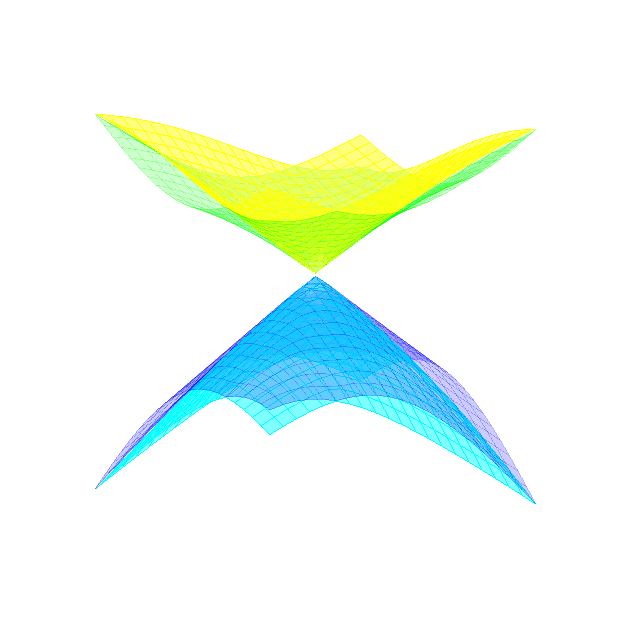}}
    \subfigure[]{\includegraphics[width=4cm]{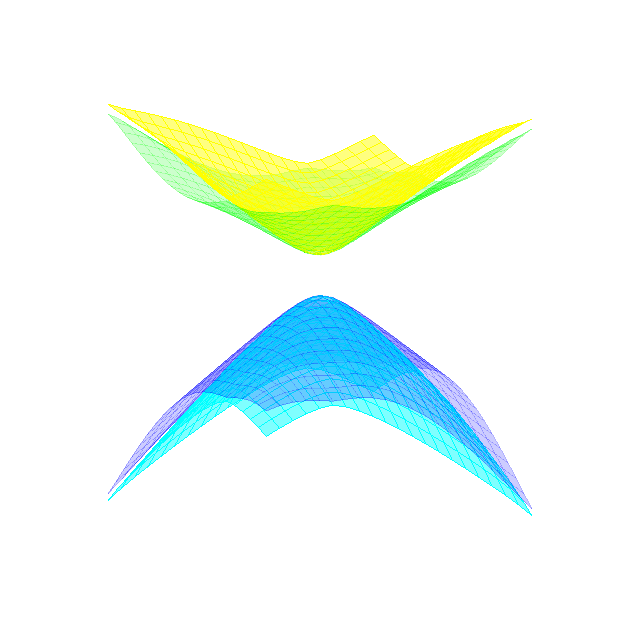}}\\
    \subfigure[]{\includegraphics[width = 4cm]{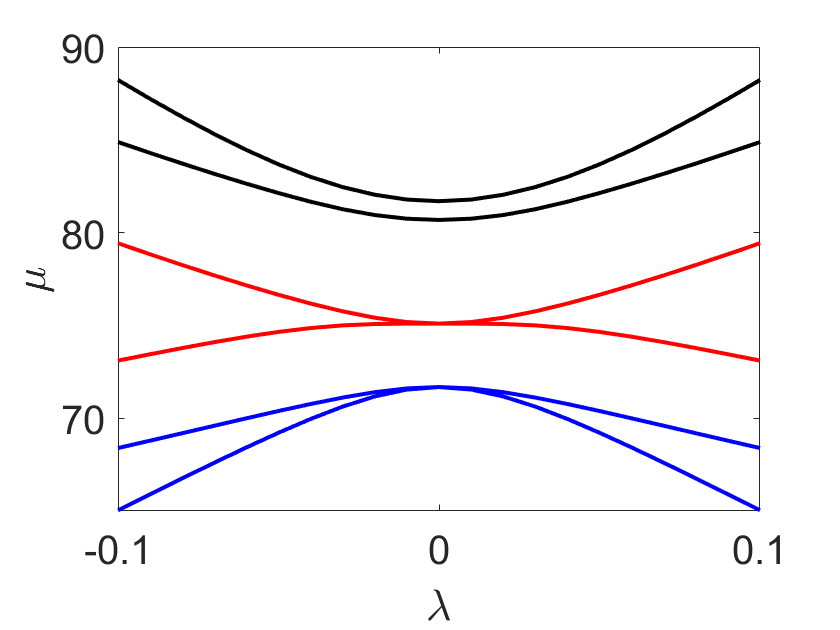}}
    \subfigure[]{\includegraphics[width = 4cm]{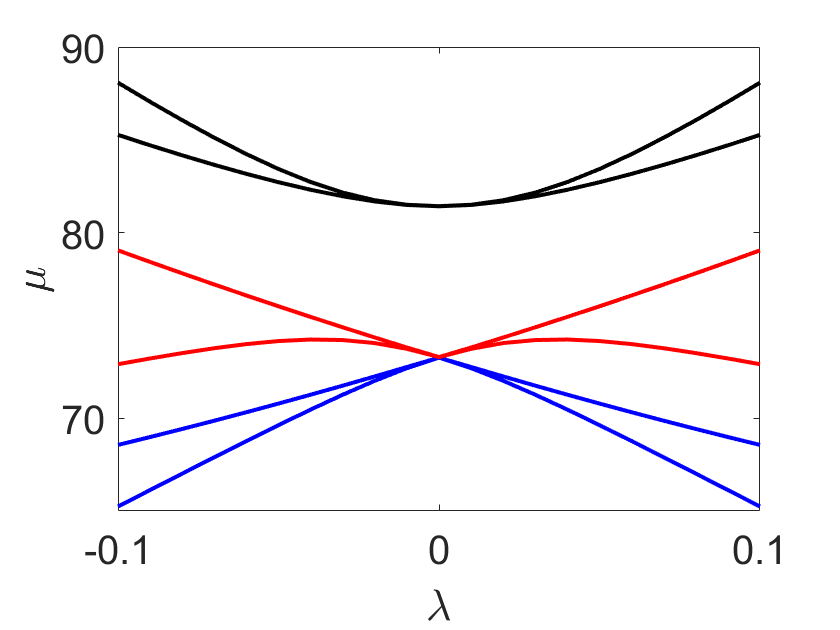}}
    \subfigure[]{\includegraphics[width = 4cm]{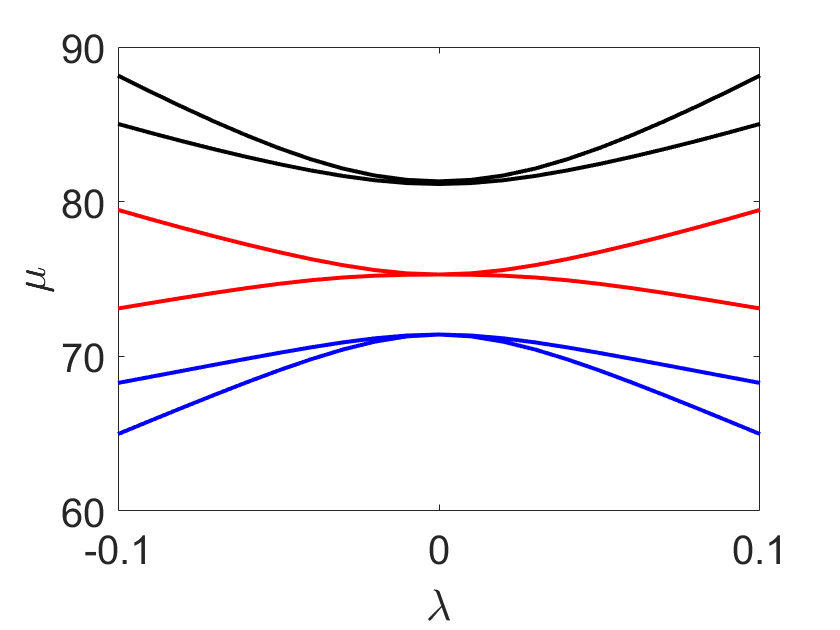}}
    \caption{Top: honeycomb dimer potentials $W(\bx,r)$ defined by (\ref{eqn-nume-f}), (\ref{eqn-dimer-g}), and (\ref{eqn-dimer-W}) with $r$ taking value (a) $\frac{1.05}{3}$, (b) $\frac{1}{3}$, and (c) $\frac{0.975}{3}$. We paint three periods in both $\buu_1$ and $\buu_2$ directions. Middle: corresponding $2^{nd}-5^{th}$ energy surfaces of $H(r)= -\Delta + W(\bx,r )$.
    Bottom: corresponding $2^{nd}-7^{th}$ energy bands along $\lambda\bk_1$.} 
    \label{figure-discontinuous}
\end{figure}

Let us observe small perturbations 
which break this additional translation symmetry as in Figure \ref{figure-perturbation}.
Shrinking and expanding the hexagons in super honeycomb lattice obtain the new lattices with red vertices.
These perturbed lattices are not super honeycomb lattices any longer, and the four branches of energy bands intersecting under super honeycomb lattice potentials' cases separate into two parts as in Figure \ref{figure-discontinuous}.

Generally, let $W(\bx)$ be a bounded real function that can be written in the form
\begin{equation}\label{eqn-W}
    W(\bx)=e^{i\bk_1\cdot \bx}p(\bx) + e^{-i\bk_1\cdot \bx}p(-\bx),\q e^{i\bk_1\cdot \bx}p(\bx) \in \chi_{\bk_1,1}
\end{equation}
Obviously, $W(\bx)$ is even, and it is $\mcr$-invariant.

After adding a perturbation of $W(\bx)$, the Schr\"odinger operator is
\begin{equation}\label{eqn-H-delta}
    H^{\delta} = H_V + \delta W(\bx) = -\Delta + V(\bx) + \delta W(\bx)
\end{equation}
The potential $V(\bx)+\delta W(\bx)$ here has $\mcr$-symmetry, $\mcp\mcc$-symmetry and translation symmetry. It is still a honeycomb lattice potential, but it is no more a super honeycomb lattice potential .

\begin{remark}\label{remark-topology}
Different ways to break the additional translation symmetry shown above represented by shrinking and expanding the lattices give different topological properties, which can be characterised by the Chern number. And gluing these two different topological materials together, say, the shrunk one and expanded one, by a domain wall, generates a new material with interesting edge states. This phenomenon will be researched in our forthcoming paper.
\end{remark}

\subsection{Band Structures after perturbations}
 Consider the perturbed eigenvalue problem:
\begin{equation}
    H^{\delta} \Phi(\bx) = \mu^{\delta}(\bk) \Phi(\bx), \q \bx \in \R^2
\end{equation}
\begin{equation}
    \Phi(\bx + \buu) = e^{i\bk \cdot \buu} \Phi(\bx), \q \buu \in \bf{U}
\end{equation}
Here $\bk$ is in $\Omega^*$. Again, let $\Phi(\bx) = e^{i\bk\cdot \bx}\phi(\bx)$ and denote $H^{\delta}(\bk) = -(\nabla + i\bk)\cdot(\nabla + i\bk) + V(\bx) + \delta W(\bx)$. We deduce that
\begin{equation}\label{eqn-perturbe-eigen-problem-1}
    H^{\delta}(\bk) \phi(\bx) = \mu^{\delta}(\bk) \phi(\bx), \q \bx \in \R^2,
\end{equation}
\begin{equation}\label{eqn-perturbe-eigen-problem-2}
    \phi(\bx + \buu) =  \phi(\bx), \q \buu \in \bf{U}.
\end{equation}

The following theorem shows that the fourfold degeneracy and double Dirac cone are not protected after the symmetry breaking perturbations we discuss in last subsection.

\begin{theorem}\label{thm-perturbe}
Let $V(\bx)$ be a super honeycomb lattice potential and W(\bx) be as in (\ref{eqn-W}). $H^{\delta}$ is as in (\ref{eqn-H-delta}). Assume that:
\begin{enumerate}
    \item $\mu_{_D}$ is an eigenvalue of $H_V$ on $\chi$ of multiplicity four:
    \begin{equation*}
        \mu_b^0(\bm{0}) < \mu_{_D} = \mu_{b+1}^0(\bm{0})  = \mu_{b+2}^0(\bm{0})  = \mu_{b+3}^0(\bm{0})
        = \mu_{b+4}^0(\bm{0})  < \mu_{b+5}^0(\bm{0}) ,
    \end{equation*}
    
    \item $\mu_{_D}$ is a simple eigenvalue on $\chi_{\bk_1,\tau}$ with normalized eigenfunction $\phi_1(\bx)$,
    \item $\bm{v}_{\sharp} = <\phi_1(\bx), \nabla \mcp\mcc[\phi_1](\bx)> \neq \bf{0}$,
    \item $c_{\sharp} = <\phi_1(\bx),W(\bx)\phi_3(\bx)>\neq 0.$
\end{enumerate}
Then due to Proposition \ref{thm-0-dirac-cone}, there exists double Dirac cone in the vicinity of the $\Gamma$ point for $H_V$. 
We claim that exists $\delta_0>0$, such that for all $\delta \in (-\delta_0,\delta_0)\setminus\{0\}$, the $(b+1)^{th}$ to $(b+4)^{th}$ energy bands of $H^{\delta}$ will open a gap in a small neighbourhood of the $\Gamma$ point.
\end{theorem}

\Proof
Again, employ the Lyapunov-
Schmidt reduction strategy. Without loss of generality, assume $<\phi_l(\bx),\phi_j(\bx)>=\delta_{l,j}$.

Now settle (\ref{eqn-perturbe-eigen-problem-1})-(\ref{eqn-perturbe-eigen-problem-1}). Let $\mu^{\delta}(\bk) = \mu_{_D} + \lambda$ and $\phi(\bx) = \alpha^j \phi_j(\bx) + \psi(\bx)$. $\psi(\bx) \in ker(H_V-\mu_{_D} I)^{\perp}$. Then (\ref{eqn-perturbe-eigen-problem-1}) is:
\begin{equation*}
    (H_V - \mu_{_D} I) \psi(\bx) = (2i\bk \nabla - |\bk|^2  -\delta W(\bx) + \lambda )(\alpha^j \phi_j(\bx) + \psi(\bx))
\end{equation*}
Apply the projection $\mcq_{\parallel}$ from $\chi$ to $ker(H_V-\mu_{_D} I)$ and $\mcq_{\perp}$ from $\chi$ to $ker(H_V-\mu_{_D} I)^{\perp}$ to obtain the equation:
\begin{equation}\label{eqn-w-parallel}
    \mcq_{\parallel}(\delta W(\bx) - 2ik\nabla ) \psi(\bx) = \mcq_{\parallel}(2i\bk \nabla - |\bk|^2  -\delta W(\bx) + \lambda )\alpha^j \phi_j(\bx),
\end{equation}
and
\begin{equation}\label{eqn-w-perp}
\begin{aligned}
    (H_V - \mu_{_D} I) \psi(\bx) = &  \mcq_{\perp}(2i\bk \nabla  -\delta W(\bx)  )\alpha^j \phi_j(\bx) \\ & +\mcq_{\perp}(2i\bk \nabla - |\bk|^2  -\delta W(\bx) + \lambda )\psi(\bx).
\end{aligned}
\end{equation}
After the analogous procedure to Proposition \ref{thm-0-dirac-cone}, Let
$$\mca =\big{(}I-(H_V-\mu_{_D} I)^{-1}\mcq_{\perp}(2i\bk \nabla - |\bk|^2  -\delta W(\bx) + \lambda) \big{)}^{-1}, $$
$$ \mct_j[\bk,\delta,\lambda](\bx) = \mca(H_V-\mu_{_D})^{-1}\mcq_{\perp}(2i\bk \nabla  -\delta W(\bx)  )\phi_j(\bx).
$$
These operators exist and $ \mct_j$ a bounded operator from $\R^3$ to $H^1_{per}$ of order $O(|\bk|+|\delta|)$ for ($|\bk|+|\delta|+|\lambda|$) small enough. Now $\psi(\bx) = \alpha^j \mct_j[\bk,\delta,\lambda](\bx) $. 
Put it into (\ref{eqn-w-parallel}), and apply inner production with $\{\phi_l(\bx)\}_l$. Finally, 
we get linear equations $M(\bk, \delta, \lambda) (\alpha^j) = 0$, where
\begin{equation*}
\begin{aligned}
     M(\bk, \delta, \lambda) =  (<\phi_l(\bx), (\lambda + 2i\bk \nabla - \delta W(\bx) - |\bk|^2)\phi_j(\bx)\\ + ( 2i\bk \nabla - \delta W(\bx))\mct_j[\bk,\delta,\lambda](\bx)>)_{l,j}.
\end{aligned}
\end{equation*}

The truncated linear part of $M(\bk, \delta, \lambda)$, or the bifurcation matrix is 
\begin{equation}
    M_0(\bk, \delta, \lambda) = (<\phi_l(\bx), (\lambda + 2i\bk\nabla - \delta W(\bx))\phi_j(\bx)>)_{l,j}
\end{equation}
The rest part is of order $O(|\bk|^2 + |\bk||\delta| + |\delta|^2)$ when $|\lambda|$ is sufficiently small:
\begin{equation}
    M_1(\bk, \delta, \lambda) = (<\phi_l(\bx), -|\bk|^2 \phi_j(\bx) +( 2i\bk \nabla - \delta W(\bx) )\mct_j[\bk,\delta,\lambda](\bx)>)_{l,j}
\end{equation}
Thus,
\begin{equation*}
    \begin{aligned}
        det(M(\bk,\delta,\lambda)) & = det (M_0(\bk,\delta,\lambda) + M_1(\bk,\delta,\lambda)) \\ &= det(M_0(\bk,\delta,\lambda)) + O((|\bk|^2 + |\bk||\delta| + |\delta|^2)^4) 
    \end{aligned}
\end{equation*}
To calculate $M_0(\bk, \delta, \lambda)$, first note that:

\begin{proposition}\label{prop-psi-w-psi}
    There exists a real number $c_{\sharp}$ such that
    \begin{equation}
        (<\phi_l(\bx), W(\bx) \phi_j(\bx)>)_{l,j} = \begin{pmatrix}
            0 & 0 & c_{\sharp} & 0 \\ 
            0 & 0 & 0 & c_{\sharp} \\
            c_{\sharp} & 0 &0 & 0 \\
            0 & c_{\sharp} & 0 & 0 \\
        \end{pmatrix}.
    \end{equation}
\end{proposition}

\Proof
$\mcp$, $\mcc$ and $\mcr$ are unit transformations. Because $W(\bx)$ is even , real and $\mcr$-invariant, it is true that for any $f$, $g$ $\in \chi$:
\begin{equation}\label{eqn-p-p}
\begin{aligned}
     <f(\bx), W(\bx)g(\bx)> & = <\mcp f(\bx),\mcp (W(\bx)g(\bx)))>\\ & = <\mcp f(\bx), W(\bx)\mcp g(\bx))>, 
\end{aligned}
\end{equation}
\begin{equation}\label{eqn-r-r}
\begin{aligned}
     <f(\bx), W(\bx)g(\bx)> & = <\mcr f(\bx),\mcr (W(\bx)g(\bx)))> \\ & = <\mcr f(\bx), W(\bx)\mcr g(\bx))>.
\end{aligned}
\end{equation}
First use (\ref{eqn-p-p}) to obtain $c_{\sharp}$ is real:
$$\overline{c_{\sharp}} = \overline{<\phi_1(\bx), W(\bx)\phi_3(\bx)>} = <\phi_3(\bx), W(\bx)\phi_1(\bx)> = <\phi_1(\bx), W(\bx)\phi_3(\bx)> = c_{\sharp}.$$
In addition,
\begin{equation*}
    \begin{aligned}
        <\phi_2(\bx), W(\bx) \phi_4(\bx)> & = <\mcc \phi_3(\bx), \mcc (W(\bx)\phi_1(\bx))> = \overline{c_{\sharp}} \\ & = <\phi_4(\bx), W(\bx) \phi_2(\bx)> = c_{\sharp}.
    \end{aligned}
\end{equation*}
Then take $f(\bx) = \phi_1(\bx)$ and $g(\bx) = \phi_2(\bx)$. Since $\phi_1(\bx)$ and $\phi_2(\bx)$ are eigenfunctions with different eigenvalues for 
$\mcr$, $<\phi_1(\bx), W(\bx)\phi_2(\bx)> = 0$. The same for $<\phi_3(\bx), W(\bx)\phi_1(\bx)>$, $<\phi_2(\bx), W(\bx)\phi_4(\bx)>$ and $<\phi_4(\bx), W(\bx)\phi_2(\bx)>$.
Then taking $f(\bx) = g(\bx) = \phi_1(\bx)$, using (\ref{eqn-W}), due to the orthogonality of $\chi_s$, $\chi_{\bk_1}$ and $\chi_{-\bk_1}$, we have
$$ <\phi_1(\bx), W(\bx)\phi_1(\bx)>= <\phi_1(\bx), (e^{i\bk_1\cdot \bx}p(\bx) + e^{-i\bk_1\cdot \bx}p(-\bx))\phi_1(\bx)> = 0.
$$
It is the same that $<\phi_l(\bx),W(\bx)\phi_l(\bx)>=0$ for all $l$.
Taking advantage of the discussion above, it is trivial to verify the result.
\qed

Apply the equations (\ref{eqn-elment-in-birfucation}) and the proposition above to get the bifurcation matrix after perturbations:
\begin{equation}
    M_0(\bk,\delta, \lambda) =  
    \begin{pmatrix}
         \lambda & 2i\bk \cdot \bm{v}_{\sharp}  & -\delta c_{\sharp} & 0   \\
         \overline{2i\bk \cdot \bm{v}_{\sharp}} &  \lambda &  0 & -\delta c_{\sharp}  \\
         -\delta c_{\sharp} & 0 &   \lambda & -2i\bk \cdot \bm{v}_{\sharp} \\
         0 & -\delta c_{\sharp} & -\overline{2i\bk \cdot \bm{v}_{\sharp}} &    \lambda
    \end{pmatrix}
\end{equation}

Thus, $det(M_0(\bk,\delta,\lambda)) =  (\lambda^2-(\delta c_{\sharp})^2 - 4|\bk \cdot \bm{v}_{\sharp}|^2)^2$. 
Now solve $det(M(\bk,\delta,\lambda)) =   (\lambda^2-(\delta c_{\sharp})^2 - 4|\bk \cdot \bm{v}_{\sharp}|^2)^2+ O((|\bk|^2 + |\bk||\delta| + |\delta|^2)^4) $. Use notation (\ref{eqn-v_F}) here.
For $\delta$ small, with $|\bk|$ small enough too, this gives four branches of this eigenvalue problem are:
\begin{equation*}
    \begin{aligned}
        \mu_{b+1}^{\delta}(\bk) = \mu_{_D} -\sqrt{(\delta|c_{\sharp} |)^2 + |v_F \cdot \bk|^2}+ O(|\delta|^2+|\delta||\bk| + |\bk|^2) , \\
        \mu_{b+2}^{\delta}(\bk) = \mu_{_D} -\sqrt{(\delta|c_{\sharp} |)^2 + |v_F \cdot \bk|^2}+ O(|\delta|^2+|\delta||\bk| + |\bk|^2) , \\
        \mu_{b+3}^{\delta}(\bk) = \mu_{_D} + \sqrt{(\delta|c_{\sharp} |)^2 + |v_F \cdot \bk|^2}+ O(|\delta|^2+|\delta||\bk| + |\bk|^2) , \\
        \mu_{b+4}^{\delta}(\bk) =\mu_{_D} + \sqrt{(\delta|c_{\sharp} |)^2 + |v_F\cdot \bk|^2}+ O(|\delta|^2+|\delta||\bk| + |\bk|^2).
    \end{aligned}
\end{equation*}

Therefore, it is obvious that if $c_{\sharp} \neq 0$, these four bands will open a gap near $\bk=\bf{0}$ for $\delta$ sufficiently small but nonzero, which 
means the fourfold degeneracy and double cone are not protected.
\qed

\section{Numerical results}\label{Num-Res}

In this section, we numerically compute the band structures for a smooth and a piecewise constant potentials to illustrate our analysis in above sections. We use the Fourier collocation
method \cite{numerical} to solve eigenvalue problems.

In our numerical simulations, we take
\begin{equation*}
    \buu_1 = \begin{pmatrix}
       \frac{\sqrt{3}}{2} \\ \frac{1}{2}
    \end{pmatrix},\qq
    \buu_2 = \begin{pmatrix}
       \frac{\sqrt{3}}{2} \\ -\frac{1}{2}
    \end{pmatrix}.
\end{equation*}
$\bv_1$ and $\bv_2$ are as in (\ref{eqn-bv}). $\bk_1$ and $\bk_2$ are dual periods of $\buu_1$ and $\buu_2$. $\bq_1$ and $\bq_2$ are dual periods of $\bv_1$ and $\bv_2$.  

The first super honeycomb lattice potential is of the form:
$$V(\bx) = \cos( \bq_1\cdot \bx) + \cos( \bq_2 \cdot \bx) + \cos( \bq_3\cdot \bx). $$
Also we introduce the perturbation, which violates the additional translation symmetry in Definition \ref{def-superHoneycomb}:
$$W(\bx) = \cos( \bk_1\cdot \bx) + \cos( \bk_2 \cdot \bx) + \cos( \bk_3\cdot \bx). $$

\begin{figure}
    \centering
    \subfigure[]{\includegraphics[width = 4cm]{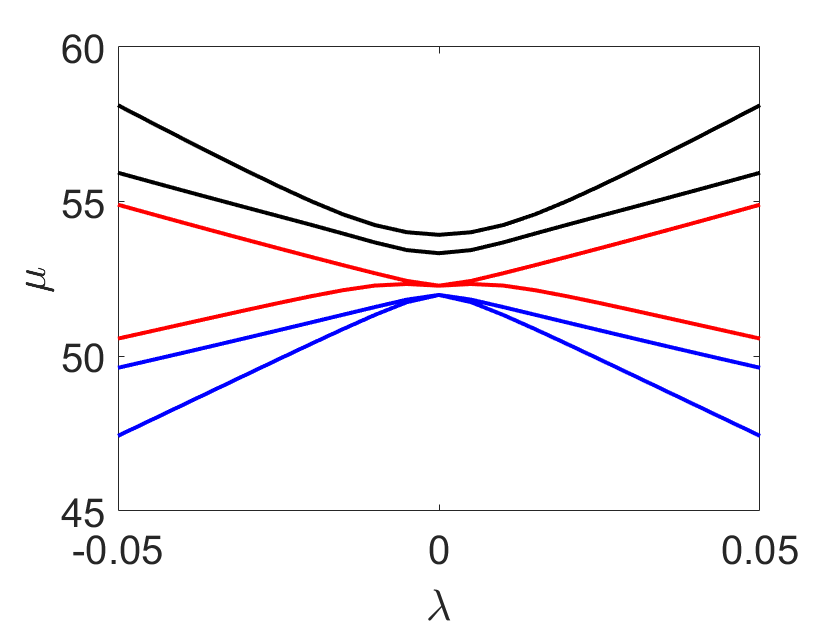}}
    \subfigure[]{\includegraphics[width=4cm]{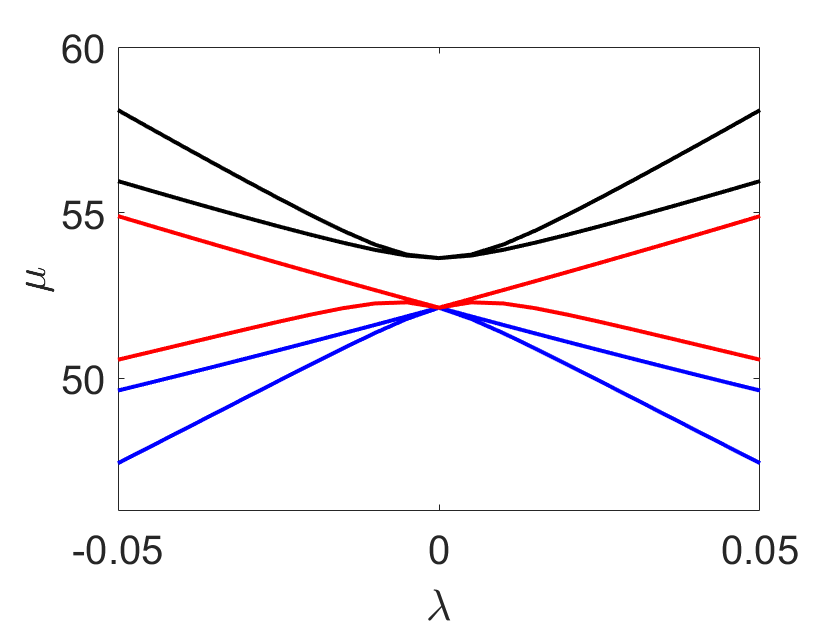}}
    \subfigure[]{\includegraphics[width=4cm]{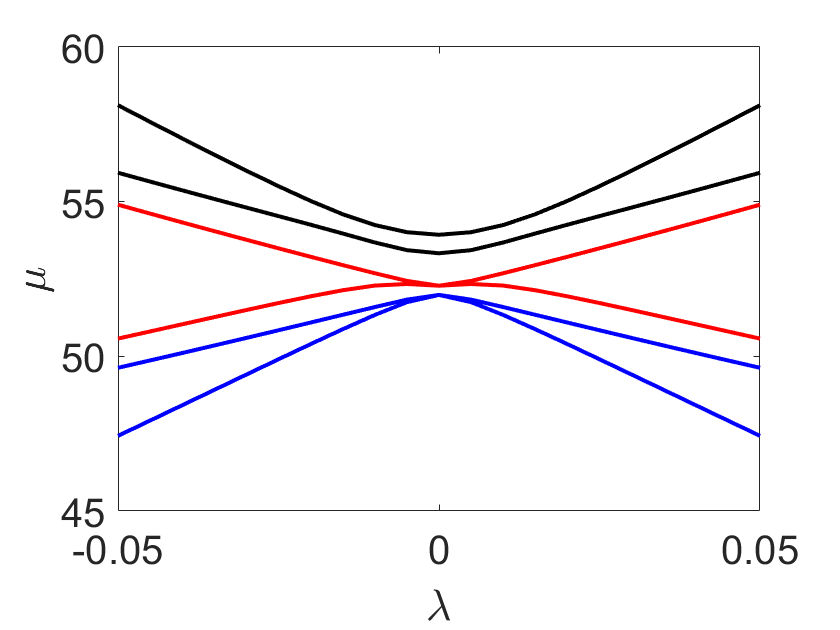}}
    \caption{The $2^{nd}$-$7^{th}$ bands for the operator $H^{\delta}(\bk) = H_V(\bk)+ \delta W(\bx)$ with $\bk = \lambda \bk_1$, where $\delta$ takes the value (a) $-0.3$, (b) $0$, and (c) $0.3$. }
    \label{figure-continuous}
\end{figure}

We compute the lowest seven bands of $H^{\delta} = -\Delta + V(\bx) + \delta  W(\bx)$ for $\delta = -0.3$, $0$, and $0.3$. The $2^{nd}-7^{th}$ bands along the $\bk_1$ direction are displayed in Figure \ref{figure-continuous}. Apparently, with the additional translation symmetry, a double Dirac cone occurs at the $\Gamma$ point. When the additional translation symmetry is broken, the fourfold degeneracy disappears and a local gap opens between the $3^{rd}$ and $4^{th}$ bands.

We construct a piecewise constant potential as follows. In the unit cell $\Omega$, take
\begin{equation}\label{eqn-nume-f}
    f(\bx) = 
    \begin{cases}
    1, & if \q \vert \bx - \frac{1}{2}(\buu_1+\buu2)\vert < 0.1  \\
    30, & else
    \end{cases}
\end{equation}
And construct $f(\bx)$ in $\R^2$ by translation along $\buu_1$ and $\buu_2$. Let $g(\bx,r)$ and $W(\bx,r)$ be in the form of (\ref{eqn-dimer-g}) and (\ref{eqn-dimer-W}). We compute the bands of $H(r) = -\Delta + W(\bx,r)$ for $r=\frac{1.05}{3}$, $r=\frac{1}{3}$, and $r=\frac{0.975}{3}$ respectively. The potentials $W(\bx,r)$ are displayed in the top panel of Figure \ref{figure-discontinuous}. The corresponding bands are displayed in the bottom panel accordingly. We remark that (b) corresponds to the super honeycomb case, i.e, possessing the additional translation symmetry. Apparently, this potential admits a fourfold degeneracy at the $\Gamma$ point and a double Dirac cone in the vicinity, but they disappear for other two cases.

\section*{Acknowledgments}
We would like to acknowledge the assistance of Borui Miao for interesting discussions and precious suggestions. 

\bibliography{references}

\end{document}